\documentclass[
 twocolumn,
 amsmath,amssymb,
 aps,
]{revtex4}

\usepackage{graphicx}
\usepackage{dcolumn}
\usepackage{bm}
\usepackage[caption=false]{subfig}
\usepackage{floatrow}
\usepackage{bm}
\usepackage[hidelinks]{hyperref}

\newcommand{\bra}[1]{\ensuremath{\langle#1|}}
\newcommand{\ket}[1]{\ensuremath{|#1\rangle}}

\newcommand{\cN}{\mathcal{N}}
\newcommand{\cC}{\mathcal{C}}
\newcommand{\cP}{\mathcal{P}}
\newcommand{\cS}{\mathcal{S}}
\newcommand{\bH}{\mathbf{H}}
\newcommand{\bx}{\mathbf{x}}
\newcommand{\bp}{\mathbf{p}}
\newcommand{\bN}{\mathbf{N}}
\newcommand{\bQ}{\mathbf{Q}}

\newcommand{\bphi}{\boldsymbol{\varphi}}
\newcommand{\pzpf}{\lambda}

\begin{document}


\title{Structurally stable subharmonic regime of a driven quantum Josephson circuit}

\author{Michiel Burgelman}
 \email{michiel.burgelman@inria.fr}
\author{Pierre Rouchon}
 \email{pierre.rouchon@minesparis.psl.eu}
\author{Alain Sarlette}
 \email{alain.sarlette@inria.fr}
\author{Mazyar Mirrahimi}%
 \email{mazyar.mirrahimi@inria.fr}
\affiliation{%
Laboratoire de Physique de l'Ecole Normale Supérieure, Inria, Mines Paris - PSL, ENS-PSL, CNRS, Sorbonne Université, Université PSL. 
}%


\begin{abstract}
Driven quantum nonlinear oscillators, while essential for quantum technologies, are generally prone to complex chaotic dynamics that fall beyond the reach of perturbative analysis. By focusing on subharmonic bifurcations of a harmonically driven oscillator, we provide a recipe for the choice of the oscillator’s parameters that ensures a regular dynamical behavior independently of the driving strength. We show that this suppression of chaotic phenomena is compatible with a strong quantum nonlinear effect reflected by the confinement rate in the degenerate manifold spanned by stable subharmonic orbits. 
\end{abstract}

\maketitle

Nonlinear oscillators are omnipresent and central for developing quantum technologies. Towards quantum computing, nonlinearity is required to realize non-Gaussian states which are a premise for quantum speedup~\cite{Lloyd1999,PhysRevLett.88.097904}. In quantum metrology, the same type of nonlinearities enable measurements of physical quantities beyond the precisions achievable with quasi-classical states~\cite{Steeneken2008,Penasa2016,Toscano2006}. In the last few decades, superconducting circuits have emerged as an exemplary platform to exploit extreme regimes of nonlinearity. Indeed, Josephson junctions coupled to microwave radiation are modelled as ideal lossless nonlinear inductors. They are now routinely used to engineer quantum states of microwave radiation, beyond what is achieved in the optical regime~\cite{Hofheinz-Nature-2009,Vlastakis2013}. The eigenstates of such anharmonic oscillators are used to encode quantum information and perform quantum operations, like logical gates and measurements, by applying appropriate driving forces~\cite{nakamura-Nature-99,PhysRevA.76.042319,Reed2010,PhysRevB.81.134507}. Furthermore, nonlinear parametric oscillation is routinely exploited to engineer processes like frequency conversion~\cite{Abdo2013}, quantum limited amplification~\cite{Castellanos-Beltran2008,Macklin2015} and multi-photon interactions~\cite{Leghtas2015}. 

All these applications have reached high precision levels where the identification of the performance limiting factors becomes extremely challenging. Indeed, while this progress has been mainly made possible due to an improved knowledge of the static properties of the system (such as the coupling to various noise sources or spurious Hamiltonians), the characterization of the impact of dynamic driving remains an outstanding problem~\cite{PhysRevLett.109.153601,PhysRevLett.109.050506,PhysRevLett.117.190503, PhysRevApplied.7.054020,Minev-Nature-2019,PhysRevX.10.011045,Leghtas2015,PhysRevX.8.021073,PhysRevA.102.022619,PhysRevApplied.11.054060,PhysRevResearch.3.033004,PhysRevApplied.11.014030,Shillito-Blais-2022}. The main approaches to address this problem have been to study microscopic effects such as drive induced quasiparticle generation~\cite{Wang-Schoelkopf-2014,PhysRevLett.113.247001}, or to develop refined perturbation theories at the macroscopic level to capture modified system parameters such as coupling to noise sources~\cite{PhysRevLett.105.100505,PhysRevLett.105.100504,PhysRevLett.118.040402,Petrescu2020,Petrescu2021,Venkatraman2021,PhysRevResearch.3.043228}. This letter puts forward another fundamental mechanism, ultimately limiting the performance of all above applications, and provides circuit design rules to prevent this from happening.

Indeed, the same driven nonlinear oscillators have  been thoroughly studied in the 1980s and 1990s from the viewpoint of complex chaotic dynamics, i.e.~observing and characterizing classical chaotic phenomena and their signature at the quantum level in driven Josephson circuits~\cite{Huberman1980,Ritala1984,Kautz_1996} and kicked rotors~\cite{casati-chirikov-79,Ammann1998}. Here, we connect these two lines of research, one on quantum information processing with superconducting nonlinear oscillators and the other, on the onset of chaotic dynamics while driving such systems. This letter hence aims to provide a recipe for selecting parameters of a quantum nonlinear oscillator where the chaotic dynamics are suppressed and one can safely rely on a refined perturbation theory to study properties of the driven system. 

\begin{figure}[h!]
\begin{center}
\includegraphics[width=\textwidth]{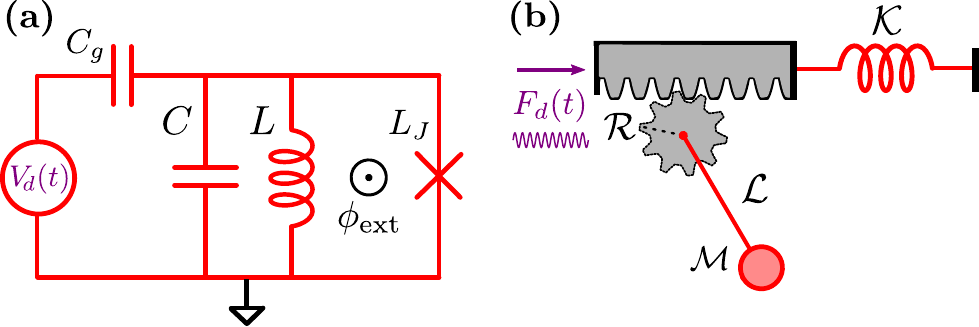}
\caption{(a) Driven  superconducting circuit comprising a Josephson junction (cross) as nonlinear element, in parallel to a linear inductance and  capacitance.
(b) Equivalent mechanical oscillator where nonlinearity is provided by the pendulum.\label{fig:circuit}}
\end{center}
\end{figure}

We consider the one-mode system represented in Fig.~\ref{fig:circuit} both as a mechanical or an electrical oscillator. This simple system is actually quite general as by changing the circuit parameters, it covers all types of superconducting qubits developed in the past decades. When varying the frequency and amplitude of a single-frequency drive  ($V_d(t)$ or $F_d(t)$ on Fig.~\ref{fig:circuit}), the classical equations of motion can undergo a subharmonic bifurcation. We discuss how this subharmonic regime mirrors a quantum parametric process confining the oscillator in a degenerate manifold of Schrödinger cat states. Next, we study the potential transition to chaos of the classical system.  We show that by appropriately  choosing one effective circuit parameter (that we call \textit{regularity parameter}), the main route to chaos is blocked in presence of any finite loss. We can therefore reliably maintain the system in the subharmonic  regime while modifying the drive parameters. We demonstrate the quantum signature of this classical transition as the breaking down of the confinement process and appearance of a high-entropy asymptotic behaviour. We next show that, with the \textit{regularity parameter} fixed to avoid such transition, a second effective circuit parameter (that we call \textit{quantum scaling parameter}) can be varied to control the quantum confinement strength in the subharmonic regime. It is thus possible to benefit from a strong nonlinear effect while maintaining the dynamics in a regular regime. While our numerical study mainly focuses on a specific subharmonic bifurcation, the results are general and apply to other subharmonics. 

The Hamiltonian of the driven circuit shown in Fig.~\ref{fig:circuit}(a) is given by
\begin{multline}\label{eq:Hamiltonian}
\bH(t)=4E_C \left(\bN-\frac{C_g V_d(t)}{2e}\right)^2+\frac{E_L}{2}\bphi^2\\-E_J\cos\left(\bphi-2\pi\frac{\phi_{\text{ext}}}{\phi_0}\right),
\end{multline}
where $E_C=e^2/2(C+C_g)$, $E_L=(\phi_0/2\pi)^2/L$, $E_J$ is the Josephson energy, $\phi_0$ is the magnetic flux quantum and $e$ is the electron charge. The operators $\bN=\bQ/2e$ and $\bphi=2e\phi/\hbar$ describe the reduced charge on the capacitance and its conjugate, the reduced  flux operator~\cite{DevoretHouches}. We take for the external magnetic flux $\phi_{\text{ext}}=0$ throughout this letter, but the methods also apply to other working points. We consider a single frequency drive $V_d(t)=\overline V_d\cos(\omega_d t)$. Rescaling $\bx= \bphi /(\sqrt{2} \pzpf)$, $\bp = \sqrt{2} \pzpf \bN$ with $\pzpf = {(2 E_C / E_L)}^{1 / 4}$, rescaling time $\tau= t \sqrt{8E_CE_L}/\hbar$ (dimensionless), and displacing the mode, the Hamiltonian becomes~\cite{Supp}:
\begin{equation}\label{eq:Hamiltonian2}
\bH(\tau)=\frac{\bp^2}{2}+\frac{\bx^2}{2}-\frac{\beta}{2 \pzpf^2}\cos(\sqrt{2}\pzpf \bx+\xi_d\sin(\nu_d\tau)).
\end{equation}
Here
\begin{align}\label{eq:params}
\beta&=\frac{E_J}{E_L}, \quad \pzpf=\left(\frac{2 E_C}{E_L}\right)^{1/4},\notag\\
\nu_d&=\frac{\hbar \omega_d}{\sqrt{8E_CE_L}},\quad \xi_d=\frac{\overline V_d C_g}{e}\sqrt{\frac{2E_C}{E_L}}\frac{\nu_d}{1-\nu_d^2}.
\end{align}
We call $\beta$ the \textit{regularity parameter} and $\pzpf$ the \textit{quantum scaling parameter} for reasons clarified below, while $\nu_d$ and $\xi_d$ are normalized parameters representing the frequency and amplitude of the drive. Finally, as relevant for quantum technologies, we consider a high-Q oscillator where energy decay is present but corresponds to the longest timescale. After a non-canonical transformation~\cite{Supp}, the corresponding classical equations of motion are:
\begin{equation}\label{eq:classical}
dx_\pzpf/d\tau= p_\pzpf,~
dp_\pzpf/d\tau=-x_\pzpf-\frac{p_\pzpf}{\tilde Q}-\beta\sin(x_\pzpf+\xi_d\sin(\nu_d \tau)).
\end{equation}
Here $1/\tilde Q$ models the decay rate with respect to dimensionless time $\tau$. Remarkably, the \textit{quantum scaling parameter} $\pzpf$ disappears from these dynamics. Note that despite the non-canonical transformation, for infinite $\tilde Q$, Eq.\eqref{eq:classical} is still of Hamiltonian form. 

\begin{figure}[h!]
\begin{center}
\includegraphics[width=\textwidth]{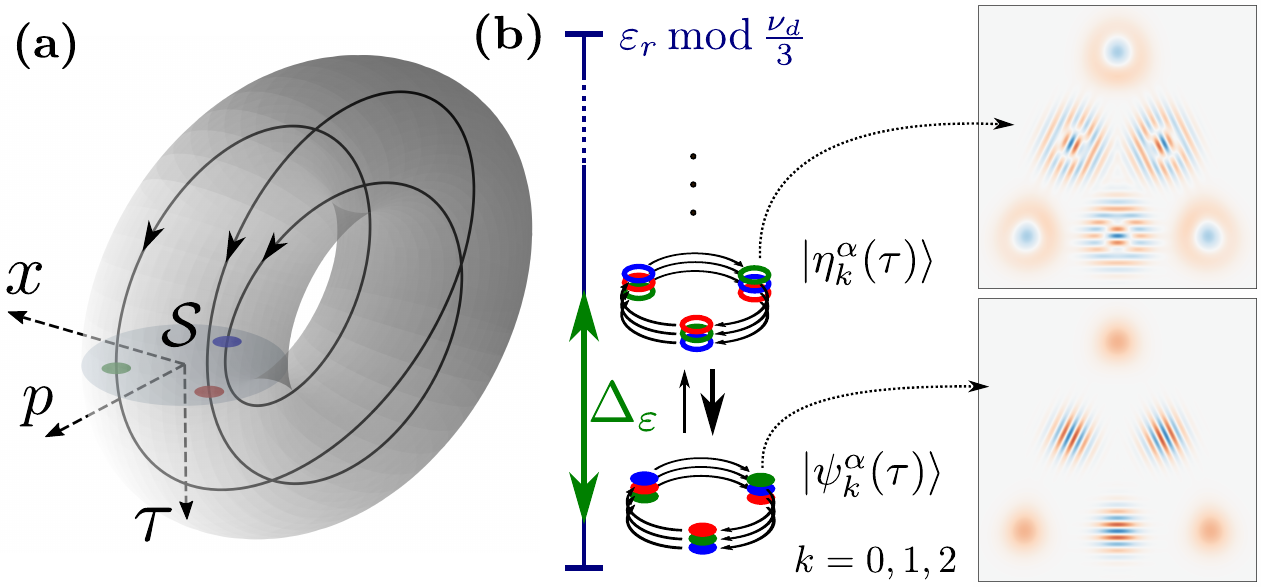}
\caption{(a) Illustration of the classical periodic planar system \eqref{eq:classical} with periodic time variable $\tau$. A (3:1) subharmonic trajectory is shown in black, intersecting the Poincaré section $\cS$ in three points (red, blue and green). (b) Schematic depiction of the equivalent quantum system~\eqref{eq:Hamiltonian2}, characterized by the Floquet modes of the driven Hamiltonian. 
The (3:1)-resonance corresponds to three dominant Floquet modes $\ket{\psi_k^\alpha(\tau)}$ with degenerate quasi-energies modulo $\nu_d/3$. Also, the three Floquet modes $\ket{\eta_k^\alpha(\tau)}$ represent the most coupled ``excited states''. The gap $\Delta_\epsilon$ between the quasi-energies of  $\ket{\eta_k^\alpha(\tau)}$ and  $\ket{\psi_k^\alpha(\tau)}$ provides the Hamiltonian confinement strength of a three component Schrödinger cat state.  \label{fig:Floquet}}
\end{center}
\end{figure}

\begin{figure*}[t!]
\begin{center}
\includegraphics[width=\textwidth]{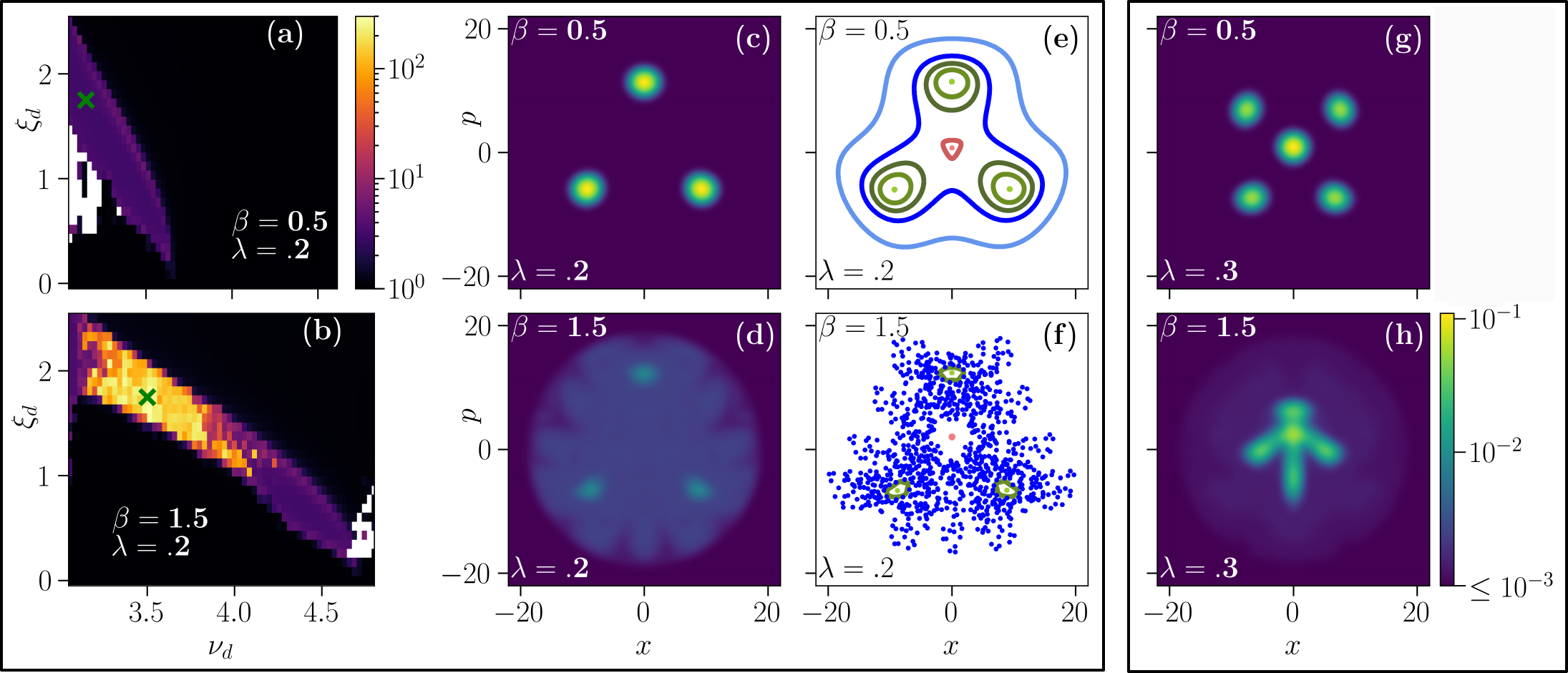}
\caption{Floquet-Markov simulations in the asymptotic regime of the weakly dissipative quantum system governed by Hamiltonian~\eqref{eq:Hamiltonian}.Plots (a)-(f) correspond to the (3:1)-resonance while the plots (g)-(h) represent a doubly degenerate (2:1)-resonance. White in plots (a) and (b) are parameter values for which the numerical simulations do not converge~\cite{Supp}.\label{fig:entropy}}
\end{center}
\end{figure*}

The time-periodic classical system \eqref{eq:classical} can be studied through the Poincaré map $\cP$ associated to  $\tau\equiv 0$(mod$2\pi/\nu_d$) (section $\cS$ in Fig.~\ref{fig:Floquet}(a)). A  fixed point of $\cP$ corresponds to a $2\pi/\nu_d$-periodic orbit of~\eqref{eq:classical}. Similarly, a subharmonic solution of period $2n\pi/\nu_d$  corresponds to $n$ fixed points of $\cP^n$. In the high-Q limit ($\tilde Q$ large), appropriate drive settings $(\nu_d,\xi_d)$ indeed permit to stabilize such subharmonic solutions, with ($n$:$m$)-resonance characterized by a period $2n\pi/\nu_d$ and a winding number $m$~\cite{Supp}. For concreteness, numerical simulations in the main text  mainly focus on the (3:1) resonance associated to the stable fixed points of $\cP^3$ (see Fig.\ref{fig:entropy}). Note that in case $n+m$ is an odd integer, a global symmetry due to the parity of the cosine potential  implies that $(n:m)$-subharmonic solutions must come in pairs~\cite{Supp}. 


The quantum dynamics can be studied through the quasi-energies and Floquet modes of the periodically driven Hamiltonian, corresponding to diagonalization of the operator $\bH(\tau)-i\partial/\partial \tau$. The Floquet mode $\ket{\Phi_{r,k}(\tau)}$ in the $k$'th Brillouin zone has quasi-energy $\epsilon_{r,k}=\epsilon_r+k\nu_d$  and satisfies $\ket{\Phi_{r,k}(\tau)}=\exp(-ik\nu_d \tau)\ket{\Phi_{r,0}(\tau)}$. In the weakly dissipative regime, the system converges to a limit cycle given by a mixture of Floquet modes $\rho_\infty(\tau)=\sum_r p_r\ket{\Phi_{r,0}(\tau)}\bra{\Phi_{r,0}(\tau)}$, where the probability distribution $\{p_r\}$ is calculated through the Floquet-Markov approach~\cite{Grifoni1998,Verney}.  The quantum mechanical counterpart of an ($n$:$m$)-resonant classical subharmonic regime, is a limit cycle mainly populated by $n$ Floquet modes $\{\ket{\Phi_{r_l,0}}\}_{l=1}^n$ with degenerate quasi-energies modulo $\nu_d/n$.
The case of a ($3$:$1$)-resonance is sketched in Fig \ref{fig:Floquet}(b), along with a Wigner function snapshot at time $\tau\equiv 0$ (mod $2\pi/\nu_d$) of associated Floquet modes. 

In other words, the above classical subharmonic regime corresponds at the quantum level to a multi-photon process confining the dynamics to a degenerate manifold of Schrödinger cat states (superpositions of distinguishable states in  phase space  that are close to coherent states). Noting again the parity of the cosine Hamiltonian, an ($n$:$m$)-resonance (where we  define $r\equiv (n+m)$(mod 2)) corresponds to a process where $(r+1)m$ drive photons at frequency $\nu_d$ are converted to $(r+1)n$  photons at frequency $m\nu_d/n$, and the degenerate manifold is spanned by cat states $\{\ket{\psi^\alpha_{k}(\tau)}\}_{k=0}^{(1+r)n-1}$ approximately given by $\ket{\cC^\alpha_{k(\text{mod}(1+r)n)}(\tau)}=1/\cN_k\sum_{l=0}^{(1+r)n-1} e^{2i\pi l k/(1+r)n}\ket{\alpha e^{2i\pi l/(1+r)n}e^{-im\nu_d\tau/n}}$~\cite{Supp}. Here, $\cN_k$ is a normalization constant, $\ket{\zeta}$ represents the coherent state of complex amplitude $\zeta$, and $\alpha$ can be tuned through drive parameters $\nu_d,\xi_d$. Such a multi-photon confinement, demonstrated recently in~\cite{Grimm2020}, is considered as a promising approach towards protecting quantum information against perturbing Hamiltonians which are weak compared to the gap $\Delta_\epsilon$ in the quasi-energy spectrum (Fig.\ref{fig:Floquet}(b)), quantifying the confinement strength~\cite{Goto2016,Puri-2017}. We further discuss this gap and its dependence on parameters at the end of this letter. 

In Figure~\ref{fig:entropy}, we study the impact of $\beta$ on the type of asymptotic behaviour. More precisely, taking the Hamiltonian~\eqref{eq:Hamiltonian}, we add weak coupling to a bath modelled by the Hamiltonian
$\sum_\omega \hbar\omega a[\omega]^\dag a[\omega]+\hbar g[\omega](a[\omega]+a[\omega]^\dag)\bp.$ For these Floquet-Markov simulations we assume a zero temperature bath with frequency-independent coupling strength $g$. We fix $\pzpf$ and study the asymptotic regime for various values of $\beta$, while varying the drive parameters. We characterize this asymptotic behavior by the entropy of $\rho_{\infty}(\tau)$, i.e. the entropy of the distribution over Floquet modes. Plots (a) and (b) of Fig.~\ref{fig:entropy} hence represent with a color axis the effective number of modes over which $\rho_{\infty}(\tau)$ is distributed, defined as
\begin{equation}\label{eq:entropy}
\exp\left(-\sum_r p_r \ln(p_r)\right) \; .
\end{equation}
Plot~\ref{fig:entropy}(a), for $\beta=0.5$, features two zones: the black one, corresponding to a dominant
harmonic solution, and the purple one, corresponding to a dominant (3:1)-subharmonic solution. Indeed, a  Husimi-Q function of $\rho_{\infty}(0)$, for drive parameters corresponding to the green cross in Fig.~\ref{fig:entropy}(a), shows essentially an equal mixture of the three states $\ket{\psi^\alpha_k(0)}$ (Fig.~\ref{fig:entropy}(c)).

On plot~\ref{fig:entropy}(b), for $\beta=1.5$, a high entropy zone appears in yellow. In this zone the subharmonic regime is essentially lost, and $\rho_{\infty}(\tau)$ spreads over a large portion of phase space ($\sim 93\%$ of population in the blue background of plot~\ref{fig:entropy}(d)). This spreading is called  wave-packet explosion and is a quantum signature of classical chaos in the weakly dissipative regime~\cite{Carlo2005,Chirikov1988}. Stronger dissipation (with respect to the Lyapunov exponents of classical chaos) would instead induce wave-packet collapse along a classical chaotic trajectory~\cite{Carlo2005,Schack1995a}.

The plots ~\ref{fig:entropy}(e) and (f) show the Poincaré maps of the associated classical dynamics~\eqref{eq:classical} in the limit of infinite $\tilde Q$. Each color corresponds to a different orbit of the Poincaré map. In (e), we see a node close to the origin, associated to a harmonic solution, encircled by red shaded orbits. Furthermore, encircled by green shaded orbits, three fixed points (centers) of $\cP^3$ are visible as three distinct phases on a period-3 orbit of $\cP$.  Finally, the blue orbits correspond to nested invariant tori (see Fig.~\ref{fig:Floquet}(a)) enclosing the harmonic and subharmonic orbits. In plot (f), the harmonic and subharmonic solutions are still present but are enclosed in a vast chaotic region, witnessed by a single orbit in blue. For a large
yet finite $\tilde Q$ (not shown), the circular features turn into tight spirals, indicating the asymptotic stability of the limit cycles, but the chaotic regime
for $\beta=1.5$ persists. 


A pattern similar to plots ~\ref{fig:entropy}(c)-(d) can be observed for the (2:1) resonance on plots ~\ref{fig:entropy}(g)-(h) (see also~\cite{Supp}). The Husimi Q-functions are given for the drive parameters $\xi_d = 1.73, \nu_d = 2.18$ (for $\beta=.5$) and $\xi_d = 1.85, \nu_d = 2.135$ (for $\beta=1.5$). The quantum scaling parameter is $\lambda=.3$: indeed, this resonance becomes extremely narrow  for smaller values of $\lambda$ and is therefore hard to identify numerically. In plot~\ref{fig:entropy}(g), we observe a mixture over the harmonic solution (yellow spot in the center) and four pair-wise degenerate Floquet modes corresponding to four-component cat states. In plot~\ref{fig:entropy}(h), we observe again the appearance of the wavepacket explosion. Further details on this odd parity resonance ($(n+m)\equiv 1$(mod2)) are provided in~\cite{Supp}. This suggests the following general picture: for low enough values of $\beta \lesssim 0.5$, target subharmonics remain robustly stable when varying the drive amplitude and accounting for the AC Stark shift. For larger values of $\beta \gtrsim 0.5$, ramping up the drive amplitude bears the danger of inducing a highly entropic regime instead of the target resonance. Note that both settings $\beta=.5$ and $\beta=1.5$ are deep in the non-hysteretic regime~\cite{Likharev1986}, since for $\phi_{\text{ext}}=0$, the potential admits no local minima for $\beta\lesssim 4.6$. The next paragraphs propose an analysis compatible with these observations. 

\begin{figure}[h!]
\begin{center}
\includegraphics[width=\textwidth]{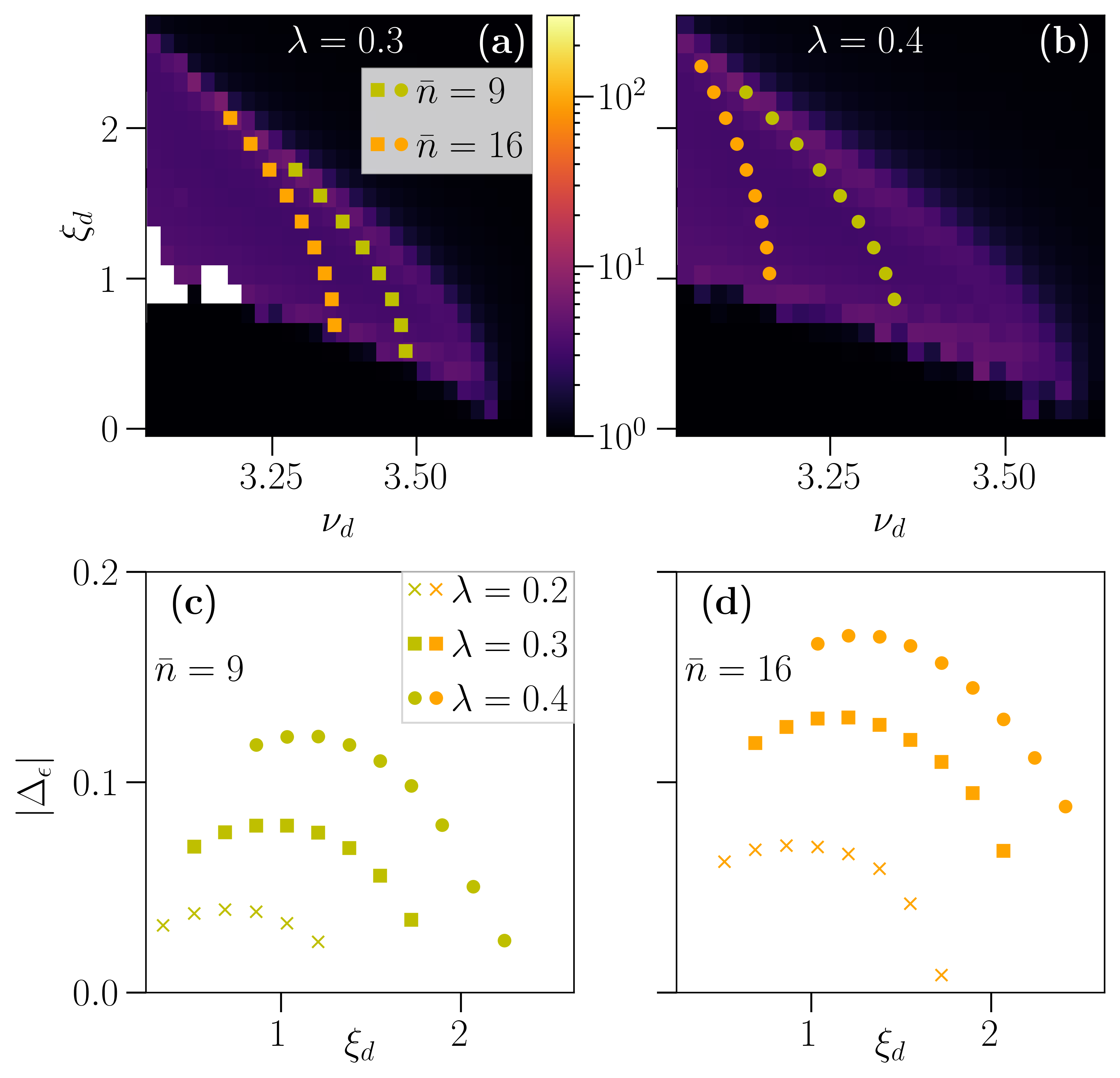}
\caption{Quasi-energy spectral gap as a function of \textit{quantum scaling parameter} $\lambda$  for fixed $\beta=0.5$ ensuring regular behaviour. \label{fig:gap}}
\end{center}
\end{figure}

In classical planar nonlinear systems with periodic drives, chaotic behavior is characterized by the presence of horseshoe dynamics~\cite{Katok1980,Smale1967}. In absence of dissipation, Hamiltonian systems can develop such structures at all spatial and temporal scales. As such, they are believed to typically feature chaotic trajectories even for very weak drives and very weak nonlinearity. For dissipative systems, there have been efforts to characterize the bifurcation mechanisms at the border of chaotic behavior.  In 1-dimensional discrete-time maps, transition to chaos always occurs through a so-called period-doubling cascade~\cite{Coilet1981,Feigenbaum1978,Feigenbaum1979,Collet1980}. It is conjectured that the same holds for 2-dimensional area-contracting, orientation-preserving embeddings of a compact disk (\cite{Gambaudo1991}, proven under additional assumptions in~\cite{Crovisier2018}). In this typical bifurcation mechanism, when a parameter is varied, a solution of a given period becomes unstable as an eigenvalue of the linearized Poincar\'e map exits the unit disk through -1; yet under the area-contracting vector field, a stable solution with double the period appears nearby. When varying the parameter further, the latter solution undergoes the same bifurcation, quadrupling the initial period. This process continues, inducing an infinite number of period doublings over a finite parameter range.


Our analytical result identifies a bound on the \emph{regularity parameter} $\beta$ to avoid the onset of such period-doubling cascade. More precisely:

\noindent \emph{Theorem:} \emph{Fix drive parameters $(\nu_d, \xi_{d})$ in Eq.\eqref{eq:classical}. Consider a stable subharmonic solution of period smaller than $\bar\tau=2\pi \bar n/\nu_d, \bar{n} \in \mathbb{N}, \bar{n} \geq 2$.
For all $\bar{\tau}$, there exists $\bar{\beta}$ such that if $0 \leq \beta<\bar\beta$,} then the solution cannot undergo a period-doubling bifurcation when varying $(\nu_d,\xi_d)$. 

This theorem, proved in~\cite{Supp}, thus excludes a period-doubling cascade starting from any such subharmonic solution. The proof provides a lower bound of $0.53/\bar \tau$ for $\bar\beta$ in the limit $\tilde Q\gg 1$.  The fact that $\bar\beta$ does not depend on $\xi_d$ results from the boundedness of the cosine potential.
All these elements together indicate that for small $\beta$, a potential chaotic regime of the classical system \eqref{eq:classical} could only originate from an extremely high-order resonance. Numerically, we have never observed a chaotic behaviour for $\beta<0.5$.

Having clarified the role of  $\beta$ in the regularity of the classical and quantum dynamics, we now focus on the other parameter $\lambda$. We remind that this parameter was suppressed in the classical dynamics, but it has an important role at the quantum level.  For fixed $\beta$ ensuring regular dynamics, we can vary $\lambda$ in the circuit design to reach strongly anharmonic quantum regimes. More precisely, developing the cosine in the Hamiltonian~\eqref{eq:Hamiltonian2} for $\xi_d=0$, the term in $\bx^4$ is proportional to $\beta\lambda^2$ while the harmonic term (in $\bx^2$) remains independent of $\lambda$. In the subharmonic  regime of the driven system, this increased anharmonicity shows up as an increased spectral gap between the degenerate manifold of cat states $\ket{\psi_k^\alpha(\tau)}$ and the next excited Floquet modes $\ket{\eta_k^\alpha(\tau)}$ (see Fig.~\ref{fig:gap}(c)-(d)). In the Kerr cat encoding~\cite{Grimm2020}, this gap equals $K|\alpha|^2$ where $K$ is the quartic Kerr strength and $|\alpha|^2 = \bar n$ the cat state's average number of photons. For our system, Figure~\ref{fig:gap} investigates different values of $\lambda$ and of the drive parameters $\xi_d,\nu_d$. Plots~\ref{fig:gap}(a) and (b) are similar to plot~\ref{fig:entropy}(a) and show that the system does not present a chaotic region for $\lambda=.2,.3,.4$. In the same plots, we mark the drive parameters leading to constant mean photon number $\bar n=9$ or $16$ in the asymptotic Schrödinger cat states. Plots (c) and (d) show the quasi-energy gap ($\Delta_\epsilon$ in Figure~\ref{fig:Floquet}(b)), corresponding to those drive parameters. Similarly to the Kerr cat case, increasing $\lambda$ ramps up the quasi-energy gap.


In conclusion, we have identified a \textit{regularity parameter} governing the structural stability  of a driven nonlinear quantum oscillator, while the quantum system can be maintained strongly  anharmonic by independently varying a \textit{quantum scaling parameter}. We characterize the loss of structural stability as a consequence of transition to chaos for the corresponding classical system, and show how a small \textit{regularity parameter} value blocks the main route towards such complex dynamics. Once regularity is ensured, we can safely rely on the \textit{quantum scaling parameter} and \textit{drive parameters}, to robustly confine  a set of subharmonic solutions, which have major applications in quantum technologies. The methods of this letter should be extendable towards multi-mode multi-drive systems required for developing large-scale quantum information devices.
\begin{acknowledgments}
The authors would like to thank Jacques Féjoz, Abed Bounemoura, Michel Devoret, Douglas Stone,  Jayameenakshi Venkatraman, Xu Xiao, Sylvain Crovisier and Christian Siegele for fruitful discussions.
This work was supported by the QuantERA grant QuCOS and by the ANR grant HAMROQS. This project has received funding from the European Research Council (ERC) under the European Union’s Horizon 2020 research and innovation programme (grant agreement No. [884762]).
\end{acknowledgments}



\bibliography{prl_refs.bib}

\end{document}



\title{Supplementary material: Structurally stable subharmonic regime of a driven quantum Josephson circuit}

\author{Michiel Burgelman}
 \email{michiel.burgelman@inria.fr}
 \author{Pierre Rouchon}
 \email{pierre.rouchon@minesparis.psl.eu}
 \author{Alain Sarlette}
 \email{alain.sarlette@inria.fr}
\author{Mazyar Mirrahimi}%
 \email{mazyar.mirrahimi@inria.fr}
\affiliation{%
Laboratoire de Physique de l'Ecole Normale Supérieure, Inria, Mines Paris - PSL, ENS-PSL, CNRS, Sorbonne Université, Université PSL.
}%


\begin{abstract}
\end{abstract}

\maketitle

\tableofcontents

\section{Normal form and classical equations of motion}\label{sec:normal_form}


In this section, we explain how the equations obtained from the quantum physical model are put into a simpler normal form towards our analytical and numerical study. We also explicitly derive the corresponding classical equations of motion.


The Hamiltonian corresponding to Figure 1.~(a) of the main text (at zero external magnetic flux) reads
\begin{equation}\label{eq:original_Hamiltonian}
\bH(t)=4E_C {\left(\bN-\frac{C_g V_d(t)}{2e}\right)}^2+\frac{E_L}{2}\bphi^2 - E_J\cos(\bphi),
\end{equation}
where $E_C$ is the charging energy, $E_L$ the inductive energy and $E_J$ is the Josephson energy.
The operators $\bN$ and $\bphi$ describe the reduced charge across the capacitance and its conjugate, the reduced flux operator.
A single-frequency drive is applied at frequency $\omega_d$:
\[V_d(t)=\overline V_d\cos(\omega_d t).\]

The  quadratic  part of the Hamiltonian~\eqref{eq:original_Hamiltonian} corresponds to a driven harmonic oscillator, and our change of variables maps it onto a normalized harmonic oscillator without drive. For this:
\begin{itemize}
    \item First, we rescale the variables ($\bphi, \bN$)$\rightarrow$ ($\tilde{\bx}= \bphi / (\sqrt{2} \pzpf)$, $\tilde{\bp} = \sqrt{2} \pzpf \, \bN$) with the \emph{quantum scaling parameter}
$\pzpf = {\qty(\frac{2 E_C}{E_L})}^{1 / 4},$ yielding
\begin{equation}
    \bH(t) = \sqrt{8 E_C E_L} \frac{\tilde\bx^2 + \tilde\bp^2}{2} - 2^{3/2} E_C \frac{C_g \bar{V}_d}{e \lambda} \cos(\omega_d t)\tilde\bp- E_J\cos(\sqrt{2} \pzpf \tilde\bx)   
\end{equation}
and maintaining the canonical commutation relations.
    \item Next, we rescale time $t$ to an undimensional time $\tau$ by multiplying it with the frequency $\sqrt{8 E_C E_L} / \hbar$, yielding
\begin{equation}
    \tilde{\bH}(\tau) = \frac{\tilde\bx^2 + \tilde\bp^2}{2} - \sqrt{\frac{E_C}{E_L}} \frac{C_g \bar{V}_d}{e \pzpf} \cos(\nu_d \tau) \tilde\bp- \frac{\beta}{2 \lambda^2}\cos(\sqrt{2} \pzpf \tilde\bx)  
\end{equation}
with 
\begin{subequations}
    \begin{align}
        \beta &=  \frac{E_J}{E_L},\\
        \tau  &=  t \frac{\sqrt{8 E_C E_L}}{\hbar},\\
        \nu_d &=  \frac{\hbar \omega_d}{\sqrt{8 E_C E_L}}
    \end{align}
\end{subequations}
\item Finally, we perform a time-dependent change of variables, following the stationary response of the linear system obtained with $\beta=0$, namely:
\begin{subequations}
\begin{align}
    \tilde{\bx} &= \bx +   \frac{\xi_d }{\sqrt{2} \pzpf} \sin(\nu_d \tau),\\
    \tilde{\bp} &= \bp + \frac{\xi_d }{\sqrt{2} \pzpf \nu_d} \cos(\nu_d \tau)\;\; , \text{ with }\\
        \xi_d       &=  \frac{\overline V_d C_g}{e}\sqrt{\frac{2E_C}{E_L}}\frac{\nu_d}{1-\nu_d^2} \; .
\end{align}
\end{subequations}
This yields
\begin{equation}\label{eq:quantum_normal_form}
    \bH(\tau)=\frac{\bp^2}{2}+\frac{\bx^2}{2}-\frac{\beta}{2 \pzpf^2}\cos(\sqrt{2} \pzpf \bx+\xi_d\sin(\nu_d\tau)) .
\end{equation}
\end{itemize}
It should be noted that~\eqref{eq:original_Hamiltonian} and~\eqref{eq:quantum_normal_form} are exactly equivalent, as no approximations have been made in the change of variables. For this reason, with a slight abuse of notation, we denote both Hamiltonians by $\bH$, the distinction being made by the time argument
$\tau$ , resp. $t$. Furthermore, we have preserved the commutation relations as $\comm{\bphi}{\bN} = \comm{\bx}{\bp} = i$. In the main text and in the following, we take \eqref{eq:quantum_normal_form} as the  quantum model to analyze.

Given their canonical commutation relations, we can consider $\bx, \bp$ as conjugate position and momentum variables $(x,\,p)$ in a corresponding classical model with Hamiltonian
$H(\tau)=\frac{p^2}{2}+\frac{x^2}{2}-\frac{\beta}{2 \pzpf^2}\cos(\sqrt{2} \pzpf x+\xi_d\sin(\nu_d\tau))$, leading to classical equations of motion (EOM)
\begin{subequations}\label{eq:unscaled_classical_system}
\begin{align}
\dv{x}{\tau} &= \frac{\partial H}{\partial p}= p,\\
\dv{p}{\tau} &= -\frac{\partial H}{\partial x}=-x-\frac{\beta}{\sqrt{2}\pzpf} \sin(\sqrt{2}\pzpf x+\xi_d\sin(\nu_d \tau))-\frac{p}{\tilde Q},
\end{align}
\end{subequations}
when accounting for a finite loss rate $1 / \tilde{Q}$ in the form of linear damping.

We can rescale the quadratures to eliminate the \emph{quantum scaling parameter} $\pzpf$ from the EOM~:
\begin{subequations}\label{eq:classical_system0}
\begin{align}
\dv{x_\pzpf}{\tau} &= p_\pzpf,\\
\dv{p_\pzpf}{\tau} &=-x_\pzpf-\frac{p_\pzpf}{\tilde Q}- \beta \sin(x_\pzpf+\xi_d\sin(\nu_d \tau)),
\end{align}
\end{subequations}
with
\begin{equation}\label{eq:AS:scaling}
    (x_\pzpf,\;p_\pzpf) = (\;\sqrt{2} \pzpf x\;\sqrt{2} \pzpf p\; ).
\end{equation}
We will study the classical equations of motion in the form \eqref{eq:classical_system0} in which the dynamics do not depend on $\pzpf$.
Note that the scaling \eqref{eq:AS:scaling} does not preserve the canonical Hamiltonian equations, so the correct quantum-classical correspondence will still involve the \emph{quantum scaling parameter} $\pzpf$.

\section{Floquet theory and numerical simulations}\label{sec:floquet_new_label}

In this section, we will review the basics of Floquet theory, both for closed Hamiltonian systems,
as for a weak coupling to a bath where the Born-Markov approximation is valid. This Floquet formalism is used to perform the numerical simulations of the present paper. 
\newcommand{\tlocal}{t}

Consider a time-dependent Hamiltonian $\bH(\tlocal)$ that is time-periodic, with period $T = 2 \pi / \omega_d$, acting on Hilbert space $\mathcal{H}$.
The Floquet theorem states that there exists solutions of the corresponding Schr\"odinger equation 
\begin{equation}\label{eq:svgl_Floquet}
    \dv{\tlocal} \ket{\psi(\tlocal)} = - \frac{i}{\hbar} \bH(\tlocal) \ket{\psi(\tlocal)},
\end{equation}
of the form
\begin{equation}
    \ket{\psi_r(\tlocal)} = e^{- \frac{i}{\hbar} \epsilon_r \tlocal} \ket{\phi_r(\tlocal)},
\end{equation}
where the \emph{Floquet modes} $\qty{\ket{\phi_r(\tlocal)}}$ form an orthonormal basis of the Hilbert space $\mathcal{H}$ at any time $\tlocal$, and are $T$-periodic:
\begin{subequations}
\begin{align}
    \ket{\phi_r(\tlocal + T)} &= \ket{\phi_r(\tlocal)}, \forall \tlocal \in \mathbb{R},\\
    \braket{\phi_r(\tlocal)}{\phi_l(\tlocal)} &= \delta_{rl}, \forall \tlocal \in \mathbb{R}.
\end{align}
\end{subequations}
Here, $\qty{\epsilon_r}$ are called the \emph{Floquet quasi-energies}. Clearly, $\epsilon_r$ is defined up to multiples of  $\hbar\omega_d$ as the Floquet modes $\ket{\phi_r(t)}$ can be multiplied by $e^{-ik\omega_d t}$. Without, loss of generality we choose the quasi-energies $\epsilon_r$ to lie in the \emph{first Brillouin zone} $[-\hbar\omega_d/2, \hbar\omega_d/2]$.
An equivalent viewpoint is that the Floquet modes (resp.~quasi-energies) are the eigenvectors (resp.~eigenvalues) of the generalized Hamiltonian
\[\bH(\tlocal) - i \hbar \pdv{\tlocal},\]
considered to act on the Hilbert space of square integrable $T$-periodic wavefunctions in $\mathcal{H}$.
Since the Floquet modes at any time $t$ form an arthonormal basis of the Hilbert space $\mathcal{H}$, one can obtain the solution of the Schr\"odinger equation corresponding
to an arbitrary initial state $\ket{\psi(0)}$ by decomposing it onto the basis of Floquet modes:
\begin{equation}
    \ket{\psi(\tlocal)} = \sum_r  e^{- \frac{i}{\hbar} \epsilon_r \tlocal} \ket{\phi_r(\tlocal)}\braket{\phi_r(0)}{\psi(0)}.
\end{equation}
The Floquet modes can be numerically computed as the eigenstates of the unitary evolution $\bU(t)$ generated by~\eqref{eq:svgl_Floquet}. Indeed, we note that the solution of the following equation
$$
\frac{d}{dt}\bU(t)=-\frac{i}{\hbar}\bH(t)\bU(t),\qquad\bU(0)=\bI
$$
with $\bI$ representing the identity operator, is given by \[\bU(\tlocal) = \sum_r  e^{ - \frac{i}{\hbar} \epsilon_r t} \ketbra{\phi_r(t)}{\phi_r(0)}.\]
Therefore, one obtains the Floquet quasi-energies and the correct basis of Floquet modes at time $\tlocal = 0$ by numerically diagonalizing $U(T)$.
Next, we can again numerically integrate the Schr\"odinger equation with initial conditions $\ket{\phi_r(0)}$ to obtain the Floquet modes at any time $t\in[0,T[$.

Floquet theory can be extended to describe the effect of a weak  coupling to a thermal bath. In particular, the asymptotic behavior of the system is described by an extension of Fermi's golden rule (\cite{Grifoni1998}, section 9). We will give a quick summary of this Floquet-Markov theory here, and explain the numerical approach that we use in the manuscript.
For concreteness, recall that in the main text,
we modeled the system coupled capacitively to a thermal bath through the Hamiltonian
\begin{equation}\label{eq:transmission_line_recap}
    \bH_{SB} = \sum_\omega \hbar \omega a[\omega]^\dag a[\omega] + \hbar g[\omega] (a[\omega] + a[\omega]^\dag) \bp,
\end{equation}
where $a[\omega]$ is the annihilation operator of the bosonic bath mode at frequency $\omega$, and $g[\omega]$
is a frequency-dependent coupling rate.
Assuming the coupling rates $g[\omega]$ to be the slowest timescale in the joint system,
and assuming a non-resonance condition on the quasi-energies (detailed here-under),
one can apply the standard Floquet-Markov-Born approximation (\cite{Grifoni1998}, section 9).
This Floquet-Markov-Born approximation yields a Lindblad type master equation for the system alone,
that can easily be solved in the Floquet basis.
When parametrizing the density matrix of the system with its components in the Floquet basis (corresponding to the first Brillouin zone),
\begin{equation}
    \rho_{rl} \coloneqq \mel{\phi_r(\tlocal)}{\rho(\tlocal)}{\phi_l(\tlocal)},
\end{equation}
one obtains a set of decoupled rate equations for $\rho_{rl}$:
\begin{subequations}\label{eq:rate_equations}
\begin{align}
    \dv{\tlocal} \rho_{rr}(\tlocal) &= \sum_l L_{rl} \rho_{ll}(\tlocal) - L_{lr} \rho_{rr}(\tlocal),\\   
    \dv{\tlocal} \rho_{rl}(\tlocal) &= - \frac{1}{2} \sum_m (L_{mr} + L_{ml}) \rho_{rl}(\tlocal).
\end{align}
\end{subequations}
The transition rates $L_{rl}$ are given by an extension of Fermi's golden rule:
\begin{equation}\label{eq:fermi_golden_rule}
    L_{rl} \coloneqq \sum_m \gamma_{r,l,m} + n_\textrm{th}\qty(\abs{\Delta_{rlm}}) \qty(\gamma_{r,l,m} + \gamma_{l,r, -m}).
\end{equation}
Here,

\begin{itemize}
\item $\hbar\Delta_{rlm} \coloneqq \epsilon_l - \epsilon_r + m \hbar \omega_d$,
\item $n_\textrm{th}(\omega)$ represents the average number of thermal photons in the bath mode at frequency $\omega$, given by the Bose-Einstein distribution
at thermal equilibrium with bath temperature $T_\textrm{bath}$:\[n_\textrm{th} = \frac{1}{\exp(\frac{\hbar \omega}{k_B T_\textrm{bath}}) - 1}.\]

\item The coefficients $\gamma_{rlm}$ are in turn given by
\begin{equation}
    \gamma_{r,l,m} \coloneqq 2 \pi \Theta(\Delta_{rlm}) J(\Delta_{rlm}) {\abs{P_{rlm}}^2},
\end{equation}
where $J(\omega)$ is the spectral function of the coupling rate to the bath,
defined in turns of the $g[\omega]$ and to be evaluated mode by mode.

\item Lastly, the matrix elements $P_{rlm}$ are defined by
\begin{equation}
    i \mel{\phi_r(\tlocal)}{a - a^\dag}{\phi_l(\tlocal)} = \sum_m P_{rlm} e^{i m \omega_d \tlocal}.
\end{equation}
\end{itemize}

The above rate equation can be contrasted to Fermi's golden rule for stationary systems as follows.
The transition frequency between two Floquet modes is now not simply given by one unique difference in
energy. Instead there are an infinite number of possible transition frequencies, shifted by  harmonics of the drive frequency $\omega_d$.
This strokes with the fact that the Floquet modes themselves can contain any  harmonic of the drive in principle.
The total transition rate between two Floquet modes $r \leftrightarrow l$ induced by the coupling to the bath is now the sum of the rates of these
different possible transitions, obtained by evaluating the bath spectral noise density at these different transition frequencies.

It is easy to see that the rate equations~\eqref{eq:rate_equations} ensure that the asymptotic density matrix $\rho_\infty(\tlocal)$ is diagonal in the
Floquet basis, since $\rho_{rl} \rightarrow 0, r \neq l$, and that the state converges to a unique classical mixture over the Floquet modes.
\[\rho(\tlocal) \rightarrow \rho_\infty(\tlocal) = \sum_r p_r \ketbra{\phi_r(\tlocal)}.\]
To compute the probability distribution $\qty{p_r}$ corresponding to this mixture, it suffices to solve the linear system of equations $Rp=0$, with
\begin{equation}\label{eq:rate_matrix}
    R_{rl} \coloneqq L_{rl} - \delta_{rl} \sum_m L_{rm}.
\end{equation}
The fact that the asymptotic density matrix is diagonal in the Floquet basis derives from a secular
approximation as part of the Floquet-Born-Markov approximation, where one neglects time-dependent terms that oscillate at frequencies $(\epsilon_r - \epsilon_l)/\hbar - m \omega_d$, whenever either $ r\neq l$ or $m \neq 0$.
We will see later on that indeed no (near-)degeneracies occur in the quasi-energy spectrum typically, so this secular approximation is justified.


In this work, we assume the limit of a cold bath, and hence assume $T_\textrm{bath} = 0$, so $n_\textrm{th} = 0$.
We furthermore assume $J(\omega) = J$ to be constant.
We observe that the maximal Brillouin-zone difference that has to be chosen in~\eqref{eq:fermi_golden_rule} for the simulations to have converged amounts to $m_\textrm{max} = 5$.

We use the Floquet toolbox of the QuTiP~\cite{Johansson2013} open source package to compute the Floquet decomposition,
using a modified version of the subroutine calculating the rate matrix in~\eqref{eq:rate_matrix}.
The number of Fock states that has to be used to simulate the system depends highly on the type of asymptotic cycle $\rho_\infty(\tau)$ obtained.
\begin{rem}\label{rem:non_convergence_new}
One numerical difficulty that occasionally presents itself lies in the numerical evaluation of the eigenvector of the transition matrix $R$ corresponding to the eigenvalue zero. Numerically,
we observe that $R$ admits at least one eigenvalue that is $0$ up to machine precision. For some system parameters the transition matrix $R$ exhibits two very-nearly degenerate eigenvectors, both corresponding to eigenvalues very close to $0$. At
this point the employed diagonalization algorithm becomes unstable, and whether the steady-state $\rho_\infty$ is uniquely defined cannot be concluded based on the numerical simulations. Such parameter values  are shown as white points in the corresponding figures in the main text. 
\end{rem}

\section{Global symmetry from parity of cosine potential} \label{ssec:parity_considerations}

The system exhibits a global symmetry, linked with the parity of the Josephson cosine potential.
We will discuss the consequences of this symmetry on the corresponding classical system, and link back to the consequences
for the quantum system at the end of this section.
This will allow us to explain the role of the parity $r = (m + n) \textrm{ mod } 2$ introduced in the main text.

Recalling the classical equations of motion~\eqref{eq:unscaled_classical_system} here,
\begin{subequations}\label{eq:recap_unscaled_classical_system}
\begin{align}
\dv{x}{\tau} &= p,\\
\dv{p}{\tau} &=-x-\frac{p}{\tilde Q}- \frac{\beta}{\sqrt{2} \pzpf} \sin(\sqrt{2} \pzpf x+\xi_d\sin(\nu_d \tau)),\label{eq:pdot_classical}
\end{align}
\end{subequations}
one can see that the transformation
\begin{equation}\label{eq:discrete_symmetry}
    \begin{cases}
        x &\rightarrow     \; \;   -   x,\\
        p &\rightarrow     \; \;   -   p,\\
        \tau &\rightarrow  \; \;   \hphantom{-} \tau + \frac{\pi}{\nu_d}.
    \end{cases}
\end{equation}
leaves~\eqref{eq:recap_unscaled_classical_system} invariant, as all individual terms change sign.
This symmetry implies that for any given solution ($x(\tau), p(\tau)$) of~\eqref{eq:recap_unscaled_classical_system},
another exact solution of the system is given by ($- x(\tau + \frac{\pi}{\nu_d}), - p(\tau + \frac{\pi}{\nu_d})$).
This observation has the consequence that when considering periodic orbits of $\pmap$, we can classify them into two distinct categories.
Consider a harmonic solution ($x_1(\tau), p_1(\tau)$) with the period of the drive, i.e.\ $\frac{2 \pi}{\nu_d}$-periodic.
As a first case, we can have that
\begin{subequations}\label{eq:discrete_symmetry_harmonic}
    \begin{align}
        x_1\qty(\tau + \frac{\pi}{\nu_d}) &= - x_1(\tau)\qq*{,} \forall \tau \in \mathbb{R},\\
        p_1\qty(\tau + \frac{\pi}{\nu_d}) &= - p_1(\tau)\qq*{,} \forall \tau \in \mathbb{R},
    \end{align}
\end{subequations}
in which case we merely obtained a particular symmetry of the given solution.
We will call harmonic solutions for which~\eqref{eq:discrete_symmetry_harmonic} is valid \emph{symmetric harmonics}.
Note that~\eqref{eq:discrete_symmetry_harmonic} immediately implies that the solution has the period of the drive,
as $x_1(\tau ) = - x_1(\tau + \frac{\pi}{\nu_d}) = {(-1)}^2 x_1(\tau + \frac{2 \pi}{\nu_d}) = x_1(\tau + \frac{2 \pi}{\nu_d})$,
and analogously for $p_1$.
As a second case, assume the orbit does not exhibit the symmetry~\eqref{eq:discrete_symmetry_harmonic},
in which case we can immediately identify a second, different solution of the system.
Both of the considered harmonics will then be called \emph{non-symmetric harmonics}.

The same reasoning can be applied to $n$-orbits of $\pmap$, i.e.~considering a solution ($x_n(\tau), p_n(\tau)$)
for which $(x_n(\tau), p_n(\tau)) = (x_n(\tau + \frac{2 n \pi}{\nu_d}), p_n(\tau + \frac{2 n \pi}{\nu_d}))$.
Indeed, applying the symmetry~\eqref{eq:discrete_symmetry} $n$ times, we have a derived symmetry of the system:
\begin{equation}\label{eq:discrete_symmetry_n}
    \begin{cases}
        x &\rightarrow     \; \;   {(-1)}^n   x,\\
        p &\rightarrow     \; \;   {(-1)}^n   p,\\
        \tau &\rightarrow  \; \;   \hphantom{-} \tau + \frac{n \pi}{\nu_d}.
    \end{cases}
\end{equation}
We can similarly classify the $n$-orbits by asking the question if
\begin{subequations}\label{eq:discrete_symmetry_subharmonic}
\begin{align}
    x_n\qty(\tau + \frac{n \pi}{\nu_d}) &= {(-1)}^n x_n(\tau),\\
    p_n\qty(\tau + \frac{n \pi}{\nu_d}) &= {(-1)}^n p_n(\tau),
\end{align}
\end{subequations}
holds or not.
We can see that when $n$ is even, this leads to a contradiction, as the period of the solution would be given by $\frac{\pi n}{\nu_d}$,
corresponding to an $n/2$-orbit of $\pmap$.
We can conclude that an $n$-orbit of $\pmap$ cannot exhibit the symmetry~\eqref{eq:discrete_symmetry_n} when $n$ is even.
However, when $n$ is odd, the symmetry is well-defined.
We then either have a so-dubbed \emph{symmetric} $n$-orbit for which~\eqref{eq:discrete_symmetry_subharmonic} is satisfied,
or we can identify a second, different $n$-orbit of $\pmap$ corresponding to ($-x_n(\tau + \frac{n \pi}{\nu_d}), -p_n(\tau + \frac{n \pi}{\nu_d})$).
The remaining question to be answered is which $n$-orbits of $\pmap$ adhere to the symmetry~\eqref{eq:discrete_symmetry_n}, and which do not.

\begin{defn}\label{defn:nm_subharmonic}
Consider a harmonic, $2 \pi / \nu_d$-periodic solution ($x_1(\tau), p_1(\tau)$) of~\eqref{eq:recap_unscaled_classical_system}.
An ($n:m$)-\emph{subharmonic} is defined as a $2 \pi n / \nu_d$-periodic solution $(x^{(n:m)}(\tau), p^{(n:m)}(\tau))$ of~\eqref{eq:recap_unscaled_classical_system}
that completes $m$ laps~\footnote{Note that two different solutions at the same time $\tau$ cannot cross, so this is well-defined.} around the harmonic solution ($x_1(\tau), p_1(\tau)$) during its period $2 n \pi/\nu_d$.
\end{defn}
In numerical simulations of the Poincar\'e map, we observe that for the specific class of ($n:m$)-subharmonics defined in Definition~\ref{defn:nm_subharmonic},  whenever $m + n$ is odd the solutions are non-symmetric, and hence always come in pairs. We can gain some intuition by anticipating some of the results of Section~\ref{chap:resonance_analysis}.
In Section~\ref{chap:resonance_analysis}, we show that for small enough values of $\beta$,
for any pair of positive integers ($m,n$) that are \emph{coprime},
there exists a prominent class of ($n:m$)-subharmonic solutions of the form
\begin{subequations}\label{eq:approximate_classical_subharmonic}
\begin{align}
    x\res(\tau) & \simeq R \sin(\frac{m}{n} \nu_d \tau + \theta),\\
    p\res(\tau) & \simeq R \cos(\frac{m}{n} \nu_d \tau + \theta),
\end{align}
\end{subequations}
for some distance $R > 0$ and angle $\theta \in [0, 2 \pi)$
that depend intricately on $\beta$ and the drive parameters ($\nu_d, \xi_d$).
We can see that when $n + m$ is odd, this solution is \emph{non-symmetric}.
Clearly, a second ($n:m$)-subharmonic resonance can readily be identified, by applying~\eqref{eq:discrete_symmetry_n} to the original solution.

An analogous discussion for the quantum system can be made.
We can make use of the following lemma to show that certain Floquet modes must necessarily come in pairs.
\begin{lem}\label{lem:doubling_floquet_modes}
Consider a Floquet mode $\ket{\phi(\tau)}$ of the quantum Hamiltonian $\bH(\tau)$ given in~\eqref{eq:quantum_normal_form} with corresponding quasienergy $\epsilon$,
\[\qty(H(\tau) - i \pdv{\tau}) \ket{\phi(\tau)} = \epsilon \ket{\phi(\tau)}.\]
Then
\[\ket{\tilde{\phi}(\tau)} \coloneqq e^{i \pi \ba^\dag \ba} \ket{\phi\qty(\tau + \frac{\pi}{\nu_d})}\]
is also a Floquet mode of $\bH(\tau)$, with the same quasienergy $\epsilon$.
\end{lem}
\begin{proof}
    Recalling
    \[ \bH(\tau)=\frac{\bp^2}{2}+\frac{\bx^2}{2}-\frac{\beta}{2 \pzpf^2}\cos(\sqrt{2} \pzpf \bx+\xi_d\sin(\nu_d\tau)), \]
    and introducing the annihilation operator $\ba = \frac{\bx + i \bp}{\sqrt{2}}$,
    it is easy to see that $\bH(\tau)$ adheres to the symmetry
    \[e^{i \pi \ba^\dag \ba} \bH\qty(\tau + \frac{\pi}{\nu_d}) = \bH(\tau) e^{i \pi \ba^\dag \ba}.\]
    This readily implies that
    \begin{align*}
        \qty(H(\tau) - i \pdv{\tau}) \ket{\tilde{\phi}(\tau)} &= \qty(H(\tau) - i \pdv{\tau}) e^{i \pi \ba^\dag \ba} \ket{\phi\qty(\tau + \frac{\pi}{\nu_d})}\\
                                                              &= e^{i \pi \ba^\dag \ba} \qty(H\qty(\tau + \frac{\pi}{\nu_d}) - i \pdv{\tau})  \ket{\phi\qty(\tau + \frac{\pi}{\nu_d})}\\
                                                              &= \epsilon e^{i \pi \ba^\dag \ba} \ket{\phi\qty(\tau + \frac{\pi}{\nu_d})}\\
                                                              &= \epsilon\ket{\tilde{\phi}(\tau)}.
    \end{align*}
\end{proof}
Lemma~\ref{lem:doubling_floquet_modes} implies that the Floquet modes must either be \emph{symmetric}, with
\begin{equation}\label{eq:symmetry_for_floquet_modes}
    e^{i \pi \ba^\dag \ba} \ket{\phi\qty(\tau + \frac{\pi}{\nu_d})} = \ket{\phi\qty(\tau)},
\end{equation}
or they must necessarily come in pairs of two \emph{non-symmetric} Floquet modes $\qty(\ket{\phi\qty(\tau)}, \ket{\tilde{\phi}\qty(\tau)})$ with 
\[\ket{\tilde{\phi}(\tau)} \coloneqq e^{i \pi \ba^\dag \ba} \ket{\phi\qty(\tau + \frac{\pi}{\nu_d})}.\]
The question is again which Floquet modes of the system are symmetric and which are not.
Using the approximate relation
\begin{equation}
    \psiktau \simeq  \ket{\mathcal{C}_{k (\!\!\textrm{ mod } n)}^{\alpha}}  \coloneqq \frac{1}{\mathcal{N}^{(n)}_k}
    \sum_{l = 0}^{n - 1} e^{2 i \frac{l k}{n} \pi} \ket{\alpha_0 e^{2 i \frac{l}{n} \pi} e^{- i \frac{m}{n} \nu_d \tau}} , k = 0,\ldots n - 1,
\end{equation}
with
\[\alpha_0 = \frac{i R e^{- i \theta}}{\sqrt{2}},\]
a completely analogous discussion as for the classical system shows that for a given pair of  integers ($n,m$), with $m + n$ being odd, the associated Floquet modes are necessarily non-symmetric and come in pairs.
This explains why for $m + n$ odd, a doubling of the number of cat-state Floquet modes was stated in the main text.

As a last point of this section, note that if we had considered a DC-biased Josephson potential
\[\beta \cos(\sqrt{2} \lambda x + \phi_\textrm{bias} + \xi_d \sin(\nu_d \tau)) \qq{,} \phi_\textrm{bias} \neq 0,\]
then the symmetry~\eqref{eq:discrete_symmetry} would be broken, as the corresponding term in the vector field (last term of~\eqref{eq:pdot_classical}) no longer exactly switches sign.
Hence, if we had taken a different flux bias point, this discrete symmetry would not have been present,
and there would be no immediate distinction between even-parity processes with $r = 0$ and odd-parity processes where $r = 1$.

\section{Analogous study of the ($2:1$)-resonance}

This section provides extra material for the case of the ($2:1$)-resonance, first illustrating the symmetry properties of this odd-parity resonance described in Section~\ref{ssec:parity_considerations},
next providing the classical picture for the chaotic case. 

\begin{figure}[ht]
\begin{center}
\includegraphics[width=0.6\textwidth]{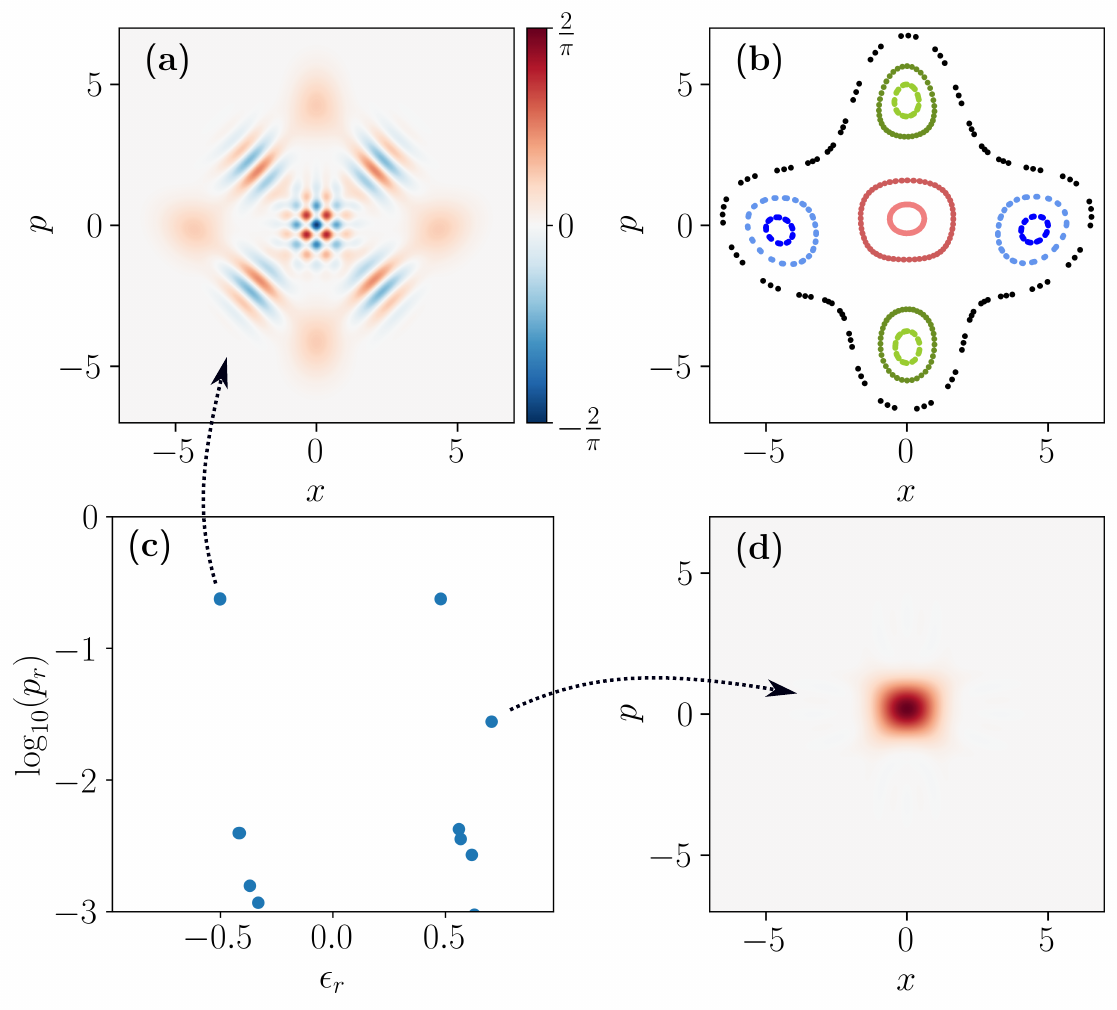}
\caption{Asymptotic behavior for the case of the ($2:1$)-resonance, for $\beta = 0.5$, $\pzpf = 0.6$, $\xi_d = 3.3$, $\nu_d = 1.96$ and $\tilde{Q} = \infty$. (a) Wigner quasiprobability representation of one of the four dominantly occupied Floquet modes $\psik$, corresponding to a four-component Schr\"odinger cat state. (b) Phase portrait of the classical Poincar\'e map $\pmap$. A stable 1-orbit of $\pmap$ is encircled by the red-shaded orbits, a stable 2-orbit is encircled by the green-shaded orbits, and a second 2-orbit is encircled by the blue orbits. (c): Quasienergies $\epsilon_r$ and occupation probabilities $p_r$ of the dominantly occupied Floquet modes in $\rho_\infty$. The four most-occupied Floquet modes are pairwise twofold degenerate, both in quasienergy $\epsilon_r$ as in occupation probability $p_r$, making them appear as only two points in total. (d) Wigner quasiprobability representation of the fifth-most occupied Floquet mode, seen to resemble a displaced vacuum state. The latter can be associated to the 1-orbit of panel (b).
\label{fig:symmetry_supmat}}
\end{center}
\end{figure}

We recall that for the classical version of the system, a ($2:1$)-resonance corresponds to two stable fixed points of $\pmap^2$, or equivalently a stable $2$-periodic orbit of 
 $\pmap$, the Poincar\'e map associated to~\eqref{eq:recap_unscaled_classical_system}. In the following we use the shorthand terminology ``2-orbit''. Due to the parity of the cosine potential, as outlined in Section~\ref{ssec:parity_considerations}, the ($2:1$)-subharmonics must come in pairs, that are related by the symmetry~\eqref{eq:discrete_symmetry_subharmonic}. An account of this can be seen in Figure~\ref{fig:symmetry_supmat}(b), where a stable 2-orbit encircled by the green orbits goes hand in hand with a stable 2-orbit encircled by the blue orbits.
The asymptotic behavior of the quantum system is represented in Figure~\ref{fig:symmetry_supmat} (a),(c),(d). In line with Lemma~\ref{lem:doubling_floquet_modes}, the most-occupied Floquet modes are indeed seen to correspond to two degenerate pairs with the same quasienergy (each pair appearing as a single point in the plot), and these Floquet modes take the form of four-component cat states (plot (a)). Unlike in the case of the ($3:1$)-resonance treated in the main text, a displaced vacuum state (plot (d)) that corresponds to a harmonic solution of the classical system shows a non-negligible occupation probability, as can be seen in plot (c). One can further discuss the dependence on the quantum scaling parameter, of the spectral gap characterizing the confinement strength in these cat states (as in Fig. 4 of the main text). Indeed, the mean number of photons in the Schrödinger cat states with the parameters of Fig.~\ref{fig:symmetry_supmat} is about 9. Changing $\lambda$ from $0.6$ to $0.4$, keeping the values of $\beta$ and $\xi_d$ the same, one can check that the same average number of photons can be reached with a choice of drive frequency $\nu_d=1.898$. Now calculating numerically the quasienergy gap as defined in the main text, one finds a value of about $0.08$ for $\lambda=0.6$ and a value of about $0.05$ for $\lambda=0.4$. This indeed supports the monotonic increase of this confinement rate with $\lambda$ as discussed in Fig.~4 of the main text.

\begin{figure}[ht]
\begin{center}
\includegraphics[width=0.9\textwidth]{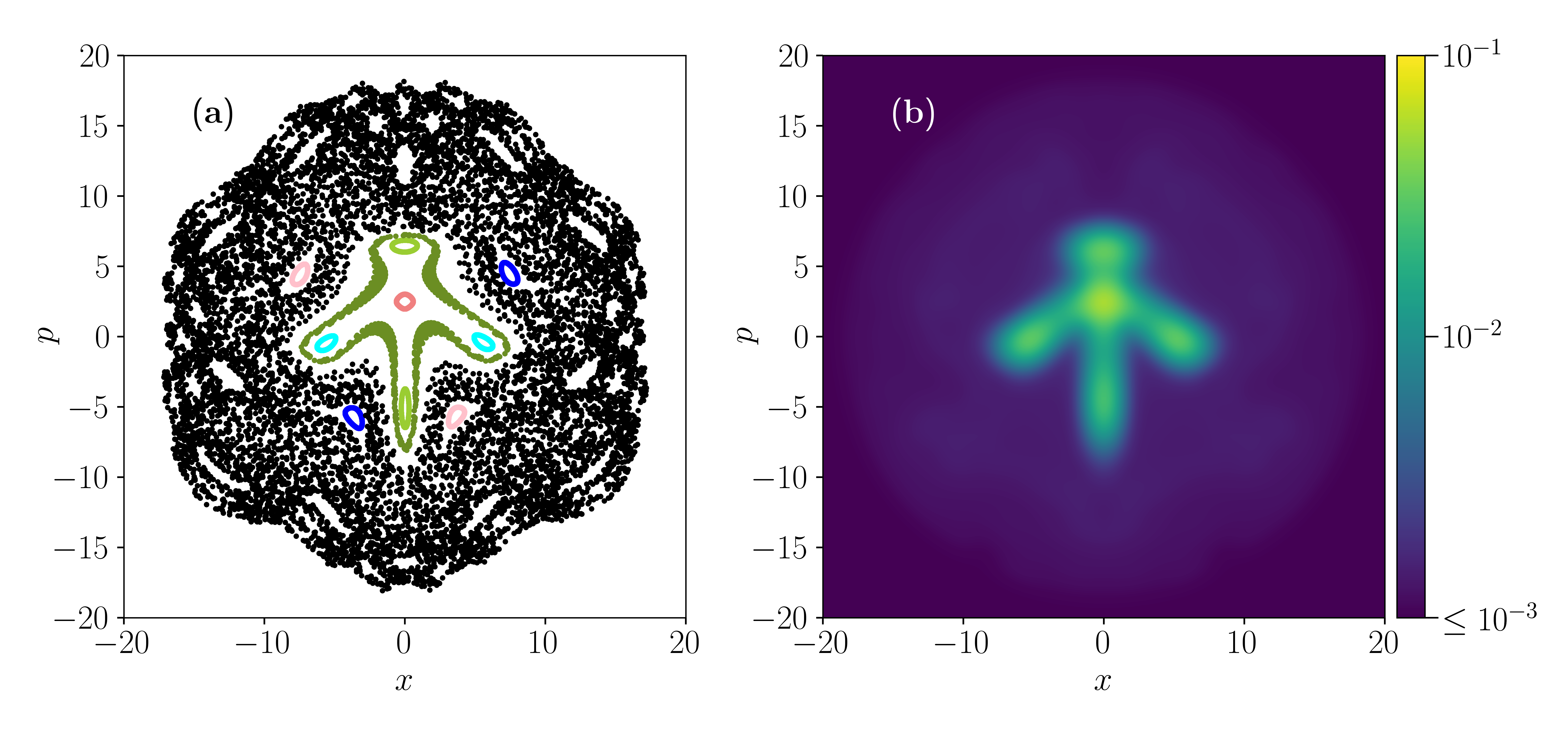}
\caption{Comparison between the asymptotic regime of the quantum and classical system, for $\beta = 1.5$, $\lambda = 0.3$, $\nu_d = 2.35$ and $\xi_d = 1.85$. This corresponds to the setting of Figure 3(h) in the main text, here repeated as plot (b) which thus shows the Husimi-Q representation of $\rho_\infty$ featuring wave-packet explosion. Plot (a) shows 7 orbits of the Poincar\'e map $\pmap$. Four different (2:1) subharmonics are observed (encircled by orbits in blue, light green, cyan, light pink), together with a chaotic orbit in dark green and a harmonic periodic orbit close to the origin (encircled by the orbit in red). Finally, a second  chaotic orbit covering a large area in phase space is shown in black. For this value of $\beta = 1.5$, many more different subharmonics coexist, albeit  with smaller domains of attraction. The extent of phase space that $\rho_\infty$ covers on plot (b) is seen to roughly correspond to the area covered by the chaotic trajectory in plot (a).
\label{fig:chaotic_explosion_supmat}}
\end{center}
\end{figure}

As a last point, while the asymptotic behavior for the case of the ($2:1$)-resonance is observed to be non-chaotic for $\beta = 0.5$, taking $\beta = 1.5$ does lead to  high-entropy asymptotic regimes characterized by $\rho_\infty$ undergoing wave-packet explosion, as was shown in Figure 3 (h) of the main text, and can be seen throughout the associated classical  Poincar\'e map in Fig.~\ref{fig:chaotic_explosion_supmat}.

\section{Avoiding chaos in the classical system}\label{sec:no_period_doubling}

In the main text, in Figure 3, we numerically show that when choosing the \emph{regularity parameter} $\beta$ small enough, we can seemingly suppress high entropy asymptotic regimes for the periodically-driven dissipative quantum nonlinear system governed by \eqref{eq:quantum_normal_form}. We further argue that such high entropy regimes of \eqref{eq:quantum_normal_form} go hand in hand with large chaotic regions in phase
space for the corresponding classical system \eqref{eq:classical_system0}. In this section, we hence study the behavior of the classical system \eqref{eq:classical_system0}, recalled here:
\begin{subequations}\label{eq:classical_system}
\begin{align}
\dv{x_\lambda}{\tau} &= p_\lambda,\\
\dv{p_\lambda}{\tau} &=-x_\lambda-\frac{p_\lambda}{\tilde Q}-\beta \sin(x_\lambda+\xi_d\sin(\nu_d \tau)),
\end{align}
\end{subequations}
and discuss how taking $\beta$ small enough would prevent it from featuring chaotic behavior.\\

\begin{figure}
    \centering
    \includegraphics[width=15cm, trim=1.9cm 6.5cm 2cm 3.5cm, clip]{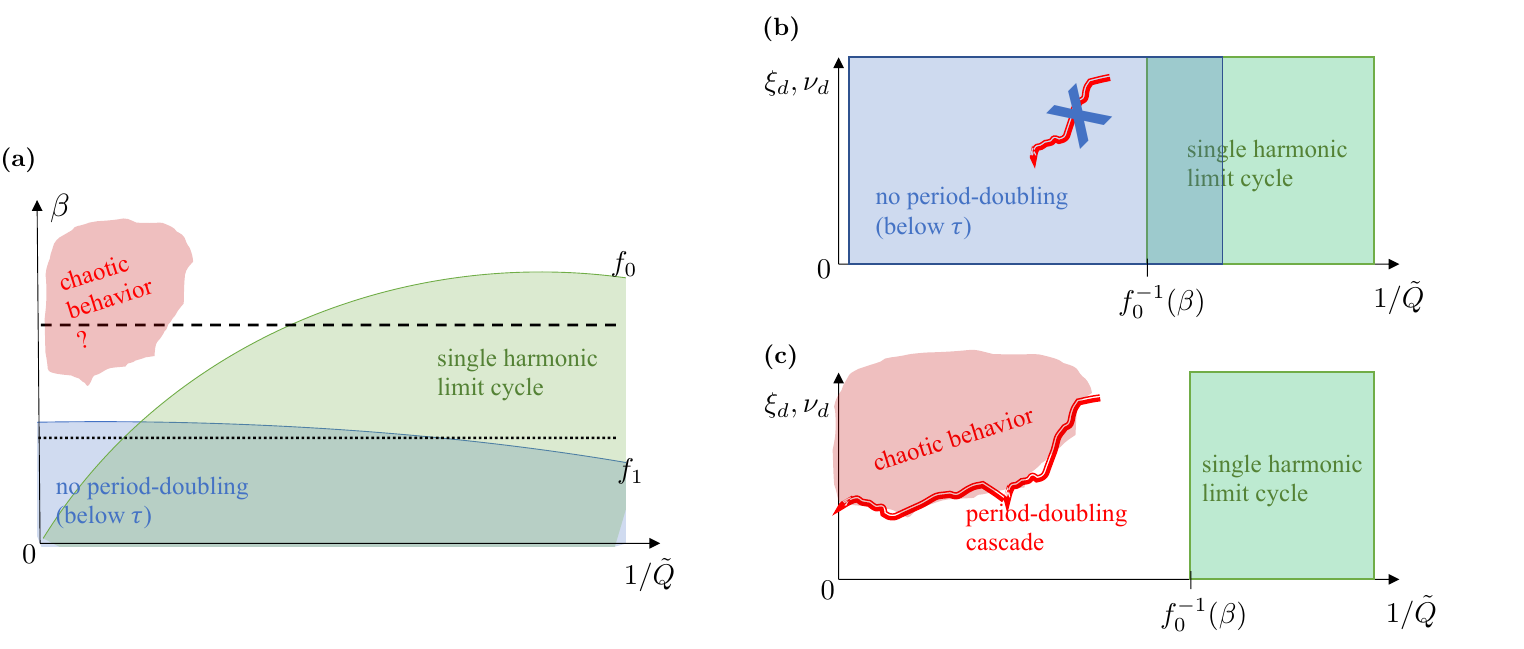}
    \caption{Schematic representation of the study about avoiding chaos in system \eqref{eq:classical_system}. \textbf{(a)} For $\beta < f_0(1/\tilde{Q})$ given in \eqref{eq:contrcondition}, corresponding to dissipation dominating nonlinearity, the system is contracting and all solutions are attracted towards a single asymptotic limit cycle (green region), whatever the drive parameters $(\xi_d,\nu_d)$. For $\beta < f_1(1/\tilde{Q},\bar{\tau})$ given by \eqref{eq:criterion}, which admits arbitrarily low dissipation, a period-doubling bifurcation is excluded, whatever the drive parameters $(\xi_d,\nu_d)$, for all solutions of period shorter than $\bar{\tau}$ (blue region). This is useful, as the Gambaudo-Tresser conjecture (here Conjecture \ref{conj:tresser}) states that if a region of parameters featuring chaotic behavior exists (red region on plots (a) and (c)), then a period-doubling cascade must exist at its boundary. \textbf{(b)} Having fixed $\beta$ at a low value, as indicated by the dotted line on plot (a), the green and blue regions as a function of the remaining parameters $(1/\tilde{Q},\xi_d,\nu_d)$ cover the whole parameter space. This excludes the existence of a period-doubling cascade, and thus of chaotic behavior for any values of $(1/\tilde{Q},\xi_d,\nu_d)$, unless the cascade is initiated by a first period-doubling of a subharmonic of very high period $>\bar{\tau}$. This absence of chaos corresponds to the situation depicted on Figure 3(e) of the main text, except now for a finite $\tilde{Q} < \infty$. \textbf{(c)} Fixing $\beta$ at a higher value, as indicated by the dashed line on plot (a), there exists a parameter zone (outside green and blue) where a period-doubling cascade can indeed be expected, such that regions featuring chaotic behavior (red zone) can exist. Such chaotic behavior is indeed observed like on Figure 3(f) of the main text, also with a finite $\tilde{Q} < \infty$.
    \label{fig:opnochaos}}
\end{figure}

The general idea is illustrated on Figure \ref{fig:opnochaos} and goes as follows.
\begin{itemize}
    \item When dissipation $1/\tilde{Q}$ dominates regularity parameter $\beta$ (green zone on Fig.~\ref{fig:opnochaos}), we can prove that the system is contracting, i.e.~any two close solutions converge towards each other, such that the asymptotic regime has to feature a single limit cycle. This is valid for any values of the drive parameters $(\xi_d,\nu_d)$, see Section \ref{ssec:app:contractive}. Our remaining goal is to extend the conclusion about the absence of chaos, from large damping $1/\tilde{Q}$ to low damping, while at the same time allowing
for additional subharmonic solutions.
    \item For appropriately fixed $\beta$, the remaining parameters $(1/\tilde{Q},\xi_d,\nu_d)$ thus feature a region where \eqref{eq:classical_system0} converges to a single limit cycle, hence there is no chaos. If there exists another parameter region featuring chaotic behavior, then it has been conjectured that the boundary to this region has to feature a period-doubling cascade -- see the Gambaudo-Tresser conjecture, here Conjecture \ref{conj:tresser} in Section \ref{ssec:app:GambaudoTresser}. 
    \item Thus conversely, by excluding the possibility of period-doubling bifurcations in a parameter region overlapping the green zone, we exclude the existence of a chaotic regime within this parameter region. For the particular system \eqref{eq:classical_system}, we prove that period-doubling is indeed impossible, at least for solutions of period smaller than $\bar{\tau}$, provided $\beta$ is small enough compared to $1/\bar{\tau}$. This criterion allows arbitrarily small damping and any values for the drive parameters (blue zone on Fig.~\ref{fig:opnochaos}); see the Theorem in the main text, recast as Corollary \ref{cor:final_no_minusone_crit} in Section \ref{ssec:app:perioddoublingbound}.
    \item All these elements together thus indicate that for low enough $\beta$, even for extremely small damping and for any values of drive parameters, the classical system should not transition into a chaotic regime. Removing two technical points would make this a rigorous result: (i) proving the Gambaudo-Tresser conjecture, which was recently done under extra technical conditions in~\cite{Crovisier2020}), and (ii), proving a bound similar to Corollary \ref{cor:final_no_minusone_crit} but independent of the period of the solution.
\end{itemize}

To simplify the mathematical analysis, in Section \ref{ssec:app:linearizing} we first perform a last change of variables on \eqref{eq:classical_system}. We also derive the linearized dynamics around a trajectory of the system, i.e.~the differential equation governing how small deviations from this trajectory will evolve over time.

\subsection{Final change of variables and local linearization around a solution}\label{ssec:app:linearizing} 

The goal of our last change of variables is to induce equal dissipation on both state variables. For this, we replace $p$ by a hyperbolically-rotated quadrature, defining
\begin{equation}\label{eq:pdiss_def}
    \tilde{p} \coloneqq \frac{p_\lambda + \frac{x_\lambda}{2 \tilde{Q}}}{\sqrt{1 - \frac{1}{4 \tilde{Q}^2}}}, \quad \xdiss = x_\lambda \; ,
\end{equation}
and rescale time as:
\begin{equation}\label{eq:AS:rescalings0}
    \tdiss \coloneqq \dissfactor\; \tau \; .
\end{equation}
Defining the modified parameters
\begin{subequations}\label{eq:AS:rescalings}
\begin{align}
    \betadiss &= \frac{\beta}{1 - \frac{1}{4 \tilde{Q}^2}},\label{eq:beta_scaling}\\
    \nuddiss &= \frac{\nu_d}{\dissfactor},\label{eq:nu_d_scaling}\\
    \kappa &= \frac{1}{2 \tilde{Q} \dissfactor},
\end{align}
\end{subequations}
we obtain the following model:
\begin{subequations}\label{eq:symmetric_dissipation}
\begin{align}
    \dv{\tdiss} \xdiss &= \hphantom{-} \pdiss - \kappa \xdiss,\\
    \dv{\tdiss} \pdiss &= - \xdiss - \kappa \pdiss - \betadiss \sin(\xdiss + \xi_d \sin(\nuddiss \tdiss)) .
\end{align}
\end{subequations}
We define the vector field $\vfdiss(\xdiss, \pdiss, \tdiss)$ such that
\begin{equation}\label{eq:vector_fieldh}
    \dv{\tdiss} \mqty(\xdiss \\ \pdiss ) = \vfdiss(\xdiss, \pdiss, \tdiss) \coloneqq \mqty(\hphantom{-} \pdiss - \kappa \xdiss\\   - \xdiss - \kappa \pdiss - \betadiss \sin(\xdiss + \xi_d \sin(\nuddiss \tdiss))),
\end{equation}
    and denote the flow corresponding to system~\eqref{eq:vector_fieldh} by $\flow_s: \mathbb{R}^2 \rightarrow \mathbb{R}^2, s \in \mathbb{R}$,
such that by definition
\begin{equation}\label{eq:EOM_flow}
    \pdv{s} \flow_s(\xdiss_0, \pdiss_0) = f(\flow_s(\xdiss_0, \pdiss_0), s).
\end{equation}
The Poincar\'e map $\pmap$, which propagates any initial condition over one period $2\pi/\nuddiss$, thus corresponds to 
$\pmap=\flow_{\perioddiss}$.\\[5mm]

We now turn to local linearization. Given any solution $(\xdiss^{(b)}(s), \pdiss^{(b)}(s)) = \flow_s(\xdiss^{(b)}_0, \pdiss^{(b)}_0)$ of the system \eqref{eq:vector_fieldh}, we can investigate how small variations $(\Delta \xdiss_0,\Delta \pdiss_0)$ of the initial condition evolve under the same dynamics, $$\flow_s(\xdiss^{(b)}_0+\Delta \xdiss,\; \pdiss^{(b)}_0+\Delta \pdiss) \;\;=\;\; (\xdiss^{(b)}(s)+\Delta \xdiss_s,\; \pdiss^{(b)}(s)+\Delta \pdiss_s) \; .$$
At the limit of infinitesimal $(\Delta \xdiss_0,\Delta \pdiss_0)$, the corresponding dynamics is given by the linearization of the vector field around the solution, i.e.:
\begin{eqnarray}\label{eq:EOM_grad_flow}
    \dv{\tdiss} \mqty(\Delta \xdiss_s \\ \Delta \pdiss_s ) & = & \grad_z f(z,s)\big\vert_{ z=(\xdiss^{(b)}(s), \pdiss^{(b)}(s))} \,  \mqty(\Delta \xdiss_s \\ \Delta \pdiss_s ) \\ & = & \nonumber
    \left(\begin{array}{ll} -\kappa & \quad 1 \\ -1 -\betadiss \cos(\xdiss^{(b)}(\tdiss) + \xi_d \sin(\nuddiss \tdiss))  & \quad -\kappa \end{array}\right) \, \mqty(\Delta \xdiss_s \\ \Delta \pdiss_s ) \; .
\end{eqnarray}
Once a solution $(\xdiss^{(b)}(s), \pdiss^{(b)}(s))$ is known, the linear time-dependent equation~\eqref{eq:EOM_grad_flow} enables the study of the system in that solution's vicinity.
We will denote  the flow corresponding to this linear system as $\Phi_s \in \mathbb{R}^{2 \times 2}$, thus satisfying
\begin{equation}
     \mqty(\Delta \xdiss_s \\ \Delta \pdiss_s ) = \Phi_s \mqty(\Delta \xdiss_0 \\ \Delta \pdiss_0),
\end{equation}
with
\begin{equation}\label{eq:AS:Phis}
    \dv{\tdiss}\Phi_s \; = \; \left(\begin{array}{ll} -\kappa & \quad 1 \\ -1 -\betadiss \cos(\xdiss^{(b)}(s) + \xi_d \sin(\nuddiss \tdiss))  & \quad -\kappa \end{array}\right) \, \Phi_s \; .
\end{equation}
By integrating over $n$ drive periods, we obtain the local linearization  of the $n$'th power of the Poincar\'e map around the chosen initial condition:
\begin{equation}
    \grad \qty(\pmap^n)(\xdiss^{(b)}_0, \pdiss^{(b)}_0) = \Phi_{2n \pi/\nuddiss}\qq{,} n \in \mathbb{N}.
\end{equation}
In particular, when ($\xdiss^{(b)}_0, \pdiss^{(b)}_0$) corresponds to a fixed point of $\pmap^n$, we can obtain the flow of the linearized system by applying a time-independent linear map:
\begin{equation}
    \grad \qty(\pmap^{n k})(\xdiss^{(b)}_0, \pdiss^{(b)}_0) =  {\qty(\grad \qty(\pmap^{n})(\xdiss^{(b)}_0, \pdiss^{(b)}_0))}^k = \Phi_{2 n k \pi/\nuddiss}\qq{,}k, n \in \mathbb{N}.
\end{equation}
%

\subsection{When damping dominates nonlinearity: single asymptotic limit cycle}\label{ssec:app:contractive}

If any two close trajectories asymptotically converge towards each other, then by induction the asymptotic regime of the system has to consist of a single trajectory. We here prove that our system  \eqref{eq:vector_fieldh} satisfies the first property, known as contraction in the dynamical systems theory, for $1/\tilde{Q}$ sufficiently large compared to $\betadiss$, and for any values of the drive parameters $(\xi_d,\nuddiss)$. We further prove that the asymptotic trajectory must be a regular $2\pi/\nuddiss$-periodic limit cycle.\\

Contraction thus analyzes local variations between two close trajectories, and therefore it studies the linearized vector field \eqref{eq:EOM_grad_flow}. In particular, contraction at a rate $r>0$ holds if we can prove that 
\begin{equation}\label{eq:AS:Contraction}
\dv{\tdiss} (\Delta x^2 + \Delta p^2) < - r \, (\Delta x^2 + \Delta p^2)\; ,
\end{equation}
for any values of $\xdiss^{(b)}(s)$ and $s$ in \eqref{eq:EOM_grad_flow}.

\begin{lem}\label{lem:contraction}
The system \eqref{eq:vector_fieldh} is a contraction, i.e.~it satisfies \eqref{eq:AS:Contraction} for a fixed $r>0$ independent of $\xdiss^{(b)}(s)$ and $s$, for any values of the drive parameters, provided
\begin{equation}\label{eq:contrcondition}
    \beta < \frac{\sqrt{1-\tfrac{1}{4 \tilde{Q}^2}}}{\tilde{Q}} \; .
\end{equation}
\end{lem}
\begin{proof}
Writing out $\dv{\tdiss} (\Delta x^2 + \Delta p^2)$ gives a quadratic expression in $(\Delta x, \Delta p)$, which satisfies \eqref{eq:AS:Contraction} provided 
$$(\kappa-r) (\Delta x^2 + \Delta p^2) + \betadiss \cos(\xdiss^{(b)}(\tdiss) + \xi_d \sin(\nuddiss \tdiss)) \Delta x \Delta p \geq 0 \; .$$
This readily gives the bound $\betadiss/2 < \kappa$, which translates into \eqref{eq:contrcondition}. 
\end{proof}

Under condition \eqref{eq:contrcondition}, any trajectory of our system is thus attracted towards a single asymptotic solution. We can further prove that this solution must be a $2\pi/\nuddiss$-periodic limit cycle. We separate the statement in two steps, as the first one will be useful in other contexts.

\begin{lem}\label{lem:diskgetsin}
Any disk around the origin of radius larger than $\betadiss/\kappa = \frac{2\beta \tilde{Q}}{\sqrt{1-\frac{1}{4\tilde{Q}^2}}}$ remains positively invariant under the evolution of the system \eqref{eq:vector_fieldh}, i.e.~when starting on this disk the trajectory moves into its interior and stays there for all times.
\end{lem}
\begin{proof}
Writing out $\dv{\tdiss} (\xdiss^2 + \pdiss^2)$,
we readily see that it is negative as soon as 
$\kappa (\xdiss^2 + \pdiss^2) > \betadiss |\pdiss|$, which holds under the stated condition on the radius.
\end{proof}

\begin{lem}\label{lem:alwayslimcycle}
For any values of the parameters, the system \eqref{eq:vector_fieldh} always features a solution that is a $2\pi/\nuddiss$-periodic trajectory.
\end{lem}
\begin{proof}
The proof uses the Brouwer fix-point theorem on the Poincar\'e map $\pmap$. This theorem states that any continuous function which maps a closed disk onto its interior, must admit at least one fix point inside this disk. The function $\pmap$ is continuous as resulting from integrating a smooth vector field, and by Lemma \ref{lem:diskgetsin} it does map any disk around the origin of radious larger than $\betadiss/\kappa$ onto its interior. Therefore $\pmap$ must always feature a fixed point, corresponding to a $2\pi/\nuddiss$-periodic trajectory of \eqref{eq:vector_fieldh}.
\end{proof}

\subsection{Linking chaos to period-doubling cascades}\label{ssec:app:GambaudoTresser}

The system \eqref{eq:vector_fieldh} is a \emph{dissipative}, periodically-driven, planar, nonlinear system. For such systems, it has been conjectured that the only possible route to chaos upon varying parameters is through a period-doubling cascade starting from an initially-stable orbit. More precisely:
\begin{conj}\emph{(Gambaudo,Tresser,~\cite{Crovisier2020,Gambaudo1991})}\label{conj:tresser}
    In the space of $\mathcal{C}^k$ orientation-preserving embeddings of a planar disk, with $k > 1$, which are area-contracting, generically,
    maps which belong to the boundary of positive topological entropy have a set of periodic orbits which,
    except for a finite subset, is made of an infinite number of periodic orbits with periods $m 2^k$ for a given $m$ and all $k \in \mathbb{N}$.
\end{conj}

The conditions of Conjecture \ref{conj:tresser} do hold for our Poincar\'e map $\pmap$ as a function of the parameters $(1/\tilde{Q},\xi_d,\nu_d)$. Indeed:
\begin{itemize}
    \item $\pmap$ is smooth for any parameter values, as resulting from the integration of a smooth vector field.
    \item $\pmap$ embeds any disk of radius larger than $\betadiss/\kappa$ into itself, as established by Lemma \ref{lem:diskgetsin}. Thus, for any fixed $\beta$ and any strictly positive interval $\tfrac{1}{\tilde{Q}} \in [\tfrac{1}{\tilde{Q}_{\max}},\tfrac{1}{\tilde{Q}_{\max}}] \subset (0, 1/ 2)$, there exists a disk of sufficient radius for which the embedding holds for all parameter values.
    \item $\pmap$ is orientation-preserving and area-contracting for any $(1/\tilde{Q},\xi_d,\nu_d)$ with $1/\tilde{Q}>0$. These are both local features to be checked uniformly in $(\xdiss, \pdiss)$ on the linearized Poincar\'e map $\grad \pmap(\xdiss, \pdiss)$. Orientation is preserved if $\det(\grad \pmap) > 0$ and area is contracted if $|\det(\grad \pmap)| < 1$. 
    From \eqref{eq:AS:Phis}, we have 
    $$\dv{\tdiss} \det(\Phi_s) = \trace\left(\begin{array}{ll} -\kappa & \quad 1 \\ -1 -\betadiss \cos(\xdiss^{(b)}(s) + \xi_d \sin(\nuddiss \tdiss))  & \quad -\kappa \end{array}\right) \; \det(\Phi_s) = -2\kappa\; \det(\Phi_s)$$
    and integrating from $\det(\Phi_0)=1$ up to $s=2\pi/\nuddiss$ we readily obtain $$\det(\grad \pmap) = \exp(- \frac{4 \pi \kappa}{\nuddiss})\; \in (0,1)\; .$$
        Physically, these properties hold by virtue of $\pmap$ being generated by a weakly dissipative system, that reduces to a Hamiltonian system for $\kappa = 0$.\\ 
\end{itemize}

We recall how this conjecture will be used in the context of our study, see Figure \ref{fig:opnochaos}(b)-(c). For fixed $\beta$, consider an open set of values for the parameters $(1/\tilde{Q},\xi_d,\nu_d)$, which we will vary to define the set of maps. In particular, the values of $1/\tilde{Q}$ should span an interval ranging from the lowest damping ever expected, up to a a value (green zone on Fig.\ref{fig:opnochaos}(b)) satisfying Lemma \ref{lem:contraction} in Section \ref{ssec:app:contractive}. Thanks to Lemma \ref{lem:contraction}, the set of parameter values thus contains settings (green zone) for which the system asymptotically converges to a single harmonic orbit. According to Conjecture \ref{conj:tresser}, if the set of parameter values also contains settings for which the system features ``positive topological entropy'', which is the technical definition of what we call ``chaos'' (red zone on Fig.\ref{fig:opnochaos}(c)), then somewhere between these two types of settings there must be a boundary with a period-doubling cascade. In the next section, to conclude our study, we will thus try to exclude the existence of a parameter region featuring chaotic behavior, by establishing conditions that exclude its boundary i.e.~period-doubling (blue zone on Fig.\ref{fig:opnochaos}(b)). 

The Gambaudo-Tresser conjecture has recently been proven under extra technical conditions in~\cite{Crovisier2020}.

\subsection{A bound on $\beta$ to exclude period-doubling}\label{ssec:app:perioddoublingbound}

\begin{figure}
    \centering
    \includegraphics[width=12cm, trim=0cm 9cm 10cm 4cm, clip]{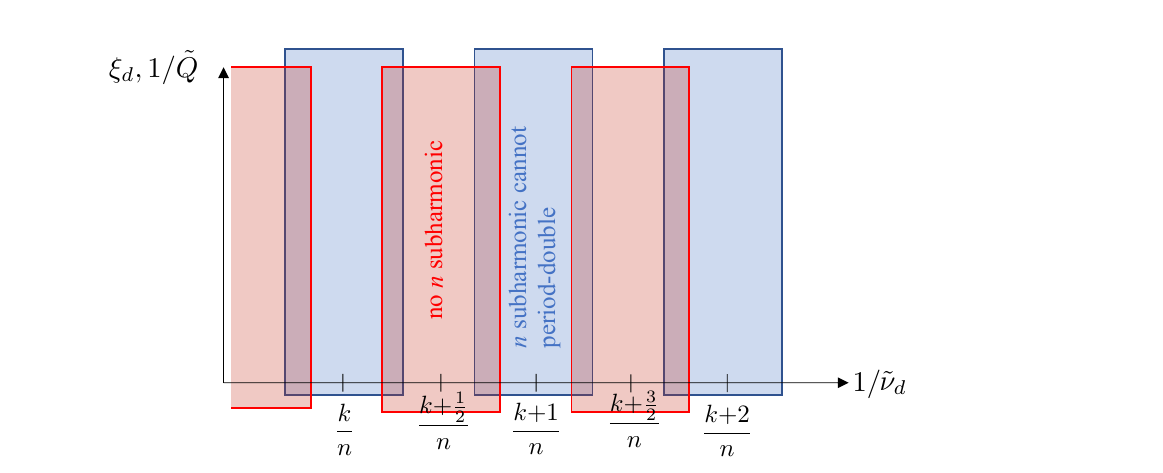}
    \caption{Splitting of the parameter space for the two parts of our proof excluding period-doubling bifurcation of a $2\pi n/\nuddiss$-periodic solution of  \eqref{eq:vector_fieldh}.}
    \label{fig:nudregions}
\end{figure}

From the previous sections, we have identified that the transition to a chaotic regime when $(\xi,\nu_d,1/\tilde{Q})$ are varied over an essentially arbitrary range, must involve the period-doubling bifurcation of periodic orbits of the system. In this last section, we establish that taking $\beta$ low enough excludes a period-doubling bifurcation for any subharmonic solution of \eqref{eq:vector_fieldh}, at least if this subharmonic's period is lower than $\bar{\tau}$, with the bound on $\beta$ depending on $\bar{\tau}$. The bound is uniformly valid for any values of $(\nu_d,\xi_d)$ and for arbitrarily low dissipation $1/\tilde{Q}$.

We thus consider as starting point a fixed value of $\bar{\tau}$ and any given subharmonic solution of \eqref{eq:vector_fieldh} of period $2\pi n/\nu_d$. We assume that $2\pi n/\nu_d < \bar{\tau}$ remains valid for any values of $\nu_d$, bounded away from zero. We then want to exclude that the considered solution undergoes a period-doubling bifurcation when varying $(\nu_d,\xi_d,1/\tilde{Q})$ in the relevant parameter range.
Our proof again works with the slightly changed coordinates~\eqref{eq:vector_fieldh} and slightly modified parameters ($\betadiss, \nuddiss$),  and it comprises two parts, see Figure~\ref{fig:nudregions}:

\begin{itemize}
    \item In Section \ref{sssec:localnobeef}, we perform a local study close to the considered subharmonic, establishing a bound on $\betadiss$ under which the premises for a period-doubling bifurcation can be excluded. This local study provides a conclusive bound only for some parameter region (blue on Fig.\ref{fig:nudregions}).
    \item In Section \ref{sssec:globalnobeef}, for another parameter region (red on Fig.\ref{fig:nudregions}), we perform a global study in phase space, showing that for low enough $\betadiss$ in fact the $2\pi n/\nuddiss$-periodic solution cannot exist.
\end{itemize}
These two parameter regions are defined as bands with $1/\nuddiss$ centered respectively around integer and half-integer multiples of $1/n$. (Recall that in \eqref{eq:vector_fieldh} the natural frequency of the harmonic oscillator for $\betadiss=0$ has been normalized to $1$.) Making these two regions overlap (Section \ref{sssec:allnobeef}), we exclude any period-doubling bifurcation of this solution within the full parameter range.\\

\subsubsection{Values of $\betadiss$ and $\nuddiss$ excluding period-doubling}\label{sssec:localnobeef}

Consider a fixed point $(\xdiss^*,\pdiss^*)$ of the smooth map $\pmap^n$, corresponding to the continuous-time trajectory $(\xdiss_n(s),
\pdiss_n(s))$ with $(\xdiss_n(0), \pdiss_n(0)) = (\xdiss^*,\pdiss^*)$, for some fixed parameter values. When varying parameters, the location and the stability of the fixed point must vary smoothly, unless it undergoes a bifurcation. A good
introduction to basic bifurcation theory can be found in~\cite{guckenheimer2002nonlinear}.
We are focusing on period-doubling bifurcations, where the initial solution becomes unstable while a stable periodic orbit of double the period
appears in its vicinity. The important property for our purposes is that at any point where $(\xdiss_n(s),
\pdiss_n(s))$ undergoes a period-doubling bifurcation, the linearized Poincar\'e map $\grad(\pmap^n)(\xdiss^*,\pdiss^*)$ must have an eigenvalue crossing $-1$~\cite{guckenheimer2002nonlinear}.
In order to exclude a period-doubling bifurcation, we thus set out to bound the
eigenvalues of $\grad \qty(\pmap^n)$ away from $-1$.
Due to the absence of an exact expression for the subharmonic solution $(\xdiss_n(s),
\pdiss_n(s))$, we approximate $\grad \qty(\pmap^n)(\xdiss^*,\pdiss^*)$ by splitting the flow $\flowdiss_s$ up into a known part, based on the linear part of the system,
and an unknown part that we treat as a perturbation, proportional to $\betadiss$. We obtain the following result.

\begin{lem}\label{lem:no_minusone_crit}
    Fix $\nuddiss >0$, and $n \in \mathbb{N}, n \geq 1$. Choose the $m \in \mathbb{N}$ which minimizes $\abs{1 - \frac{m}{n} \nuddiss}$,
    and define the detuning
    \begin{equation}\label{eq:delta_def}
        \delta := 1 - \frac{m}{n} \nuddiss.
    \end{equation}
    for which thus $|\delta| \leq \frac{\nuddiss}{2n}$.
    If
    \begin{equation}\label{eq:ASNcrit1}
        \exp(\betadiss \frac{4 \pi n}{\nuddiss}) -1 \quad < \quad 2 \cos\qty(\bar{\delta} \frac{\pi n}{\nuddiss}) \qq{, with} \abs{\delta} \leq \abs{\bar{\delta}} \leq \frac{\nuddiss}{2n}.
    \end{equation}
    then $\grad \qty(\pmap^n)(\xdiss,\pdiss)$ obtained by integrating \eqref{eq:AS:Phis} cannot exhibit an eigenvalue $-1$ for any point $(\xdiss,\pdiss) \in \mathbb{R}^2$.
    Therefore, under condition \eqref{eq:ASNcrit1}, a $2\pi n / \nuddiss$-periodic subharmonic cannot undergo a period-doubling bifurcation.
\end{lem}
\begin{proof}
    The proof is organized as follows.
    \begin{itemize}
        \item We first perform a change of variables that integrates out the time-independent part of the linearized dynamics determining, which is also independent of $\betadiss$.
        \item We then bound the spectral norm
        of the flow operator corresponding to
        this linearized dynamics uniformly in the remaining parameters. This bounds the effect of the terms proportional to $\betadiss$.
        \item From this, we next bound the spectral norm of the difference between the flow operator at time $s = 0$ (which is the identity matrix) and the flow at any time $s$ later.
        \item Finally, we use this proximity in spectral norm to deduce information about the eigenvalues of the original flow operator,
        in particular evaluating a parameter setting which guarantees that the eigenvalues cannot reach $-1$.
    \end{itemize}
    $\bullet$ \emph{Change of variables:} We start by moving to a rotating frame with frequency $\frac{m}{n} \nuddiss$, such that
    \begin{equation}\label{eq:rot_frame}
        \mqty(u(\tdiss) \\ v(\tdiss)) \coloneqq \mqty(\cos(\frac{m}{n} \nuddiss \tdiss) & - \sin(\frac{m}{n} \nuddiss \tdiss) \\ \sin(\frac{m}{n} \nuddiss \tdiss) &
        \hphantom{-} \cos(\frac{m}{n} \nuddiss \tdiss))  \mqty(\xdiss(\tdiss) \\ \pdiss(\tdiss)).
    \end{equation}
    Note that for the flow $\flowdiss_s\res$ corresponding to the variables $(u, v)$ we still have
    $\flowdiss_{T_n}\res = \flowdiss_{T_n} = \pmap^n$,
    where we introduced the total period
    \[T_n = \frac{2 \pi n}{\nuddiss},\] due to the periodicity of the change of variables, 
    Applying the change of variables to \eqref{eq:AS:Phis}, we obtain the following evolution equation for
    $\grad \flowdiss\res_\tdiss(u_0, v_0)$, the linearized flow around an arbitrary solution $(u(\tdiss), v(\tdiss))$ with $(u(0), v(0)) = (u_0, v_0)$:
    \begin{equation}\label{eq:AStointeg}
        \pdv{\tdiss} \grad \flow_\tdiss\res(u_0, v_0) = \qty(- \kappa \mqty(1 & 0 \\ 0 & 1) + \delta \mqty(0 & 1 \\ -1 & 0) + \betadiss \,
        \Gamma\res_\tdiss(u(\tdiss), v(\tdiss), \xi_d))\grad \flowdiss\res_\tdiss(u_0, v_0),
    \end{equation}
    where
    \begin{align*}
        & \Gamma\res_\tdiss(u(\tdiss), v(\tdiss),\xi_d)
        :=  \cos(\zeta(\tdiss))
        \mqty(\sin(\frac{m}{n} \nuddiss \tdiss) \cos(\frac{m}{n} \nuddiss \tdiss)   &  \sin^2(\frac{m}{n} \nuddiss \tdiss)  \\ & \\  - \cos^2(\frac{m}{n} \nuddiss \tdiss) &  -
        \sin(\frac{m}{n} \nuddiss \tdiss) \cos(\frac{m}{n} \nuddiss \tdiss)),
    \end{align*}
    with $\zeta(s) = u(\tdiss) \cos(\frac{m}{n} \nuddiss \tdiss) + v(\tdiss) \sin(\frac{m}{n} \nuddiss \tdiss) + \xi_d \sin(\nuddiss \tdiss)$.
    We will drop the reference to the particular solution $(u_0(\tdiss), v_0(\tdiss))$ for notational convenience.

    To conclude the proof we must bound the eigenvalues of $\grad{\flowdiss}_{T_n}=\grad\pmap^n$ away from $-1$. Since $\flowdiss_0$ is the identity map, $\grad \flowdiss_0 = \mathbb{1}$, where $\mathbb{1}$ is the $2 \times 2$
    identity matrix, with eigenvalues $+1$. Our strategy is to show that the eigenvalues cannot move far away from $1$ when integrating \eqref{eq:AStointeg} over a time $T_n$. We can already explicitly integrate the time-independent part, corresponding to $\beta=0$, by defining
    \begin{equation*}
        X(\tdiss) = \exp(\qty(\kappa \mathbb{1} - \delta \smqty(0 & 1 \\ -1 & 0)) \tdiss) \grad \flow_\tdiss \; ,
    \end{equation*}
    yielding
    \begin{equation}\label{eq:right_frame}
        \dv{\tdiss} X(\tdiss) = \betadiss R_{\delta \tdiss} \, \Gamma\res_s(\xi_d) \, R_{- \delta \tdiss}\; X(\tdiss),
    \end{equation}
    where $R_{\delta \tdiss}$ is the rotation matrix \[R_{\delta \tdiss} = \mqty(\cos(\delta \tdiss) & - \sin(\delta \tdiss) \\ \sin(\delta \tdiss) & \cos(\delta \tdiss)).\]

\noindent  $\bullet$ \emph{Bounding the norm of $X$:} We can write~\eqref{eq:right_frame} as an integral equation:
    \begin{equation}\label{opnieuw}
        X(s) = X(0) +  \betadiss \int_0^s R_{\delta z} \, \Gamma\res_z(\xi_d) \, R_{- \delta z} X(z) \,\dd z.
    \end{equation}
    Taking the spectral norm of both sides, applying the triangle inequality, pulling the norm into the integral in the right hand side,
    and subsequently using the submultiplicativity of the spectral norm, we obtain
    \begin{equation}\label{eq:ASbeforeGroenwall}
        \norm{X(\tdiss)} \leq \norm{X(0)}  + \betadiss \int_0^s \norm{X(z)}\, \norm{\Gamma\res_z(\xi_d)} \,\dd z.
    \end{equation}
    Since the entries of $\Gamma\res_z$ are all smaller than $1$ in absolute value, uniformly in $z$ and $\xi_d$, we have $\norm{\Gamma\res_\tdiss(\xi_d)} \leq 2$. Plugging this into \eqref{eq:ASbeforeGroenwall} and applying the simplest form of the Gr\"onwall Lemma (recalled as Lemma~\ref{gronwall_lemma} below) then yields
    \begin{equation}\label{eerste_ong}
        \norm{X(\tdiss)} \leq \norm{X(0)} e^{2 \betadiss \tdiss} = e^{2 \betadiss \tdiss},
    \end{equation}
    since $X(0) = \mathbb{1}$.

 \noindent   $\bullet$ \emph{Tying $X$ to the identity:} Now, consider again Equation~\eqref{opnieuw}:
    \begin{equation}
        X(\tdiss) - \mathbb{1} = \int_0^\tdiss \betadiss R_{\delta z} \, \Gamma\res_z(\xi_d) \, R_{- \delta z} X(z) \,\dd z.
    \end{equation}
    Analogously to the previous point, taking the spectral norm of both sides, subsequently pulling the norm into the integral into the right hand side, using the submultiplicativity of the spectral norm and the bound on $\norm{\Gamma}$, we obtain
    \begin{equation}
        \norm{X(\tdiss) -1} \leq 2 \betadiss \int_0^t \norm{X(\tdiss)}\,\dd \tdiss.
    \end{equation}
    Plugging~\eqref{eerste_ong} in the right hand side and evaluating at $s=T_n$, we obtain
    \begin{equation}
        \norm{X(T_n) -\mathbb{1}} \leq e^{2 \betadiss T_n} - 1.
    \end{equation}

\noindent $\bullet$ \emph{Confining the eigenvalues of $\grad{\flowdiss}_{T_n}$:}
    We have thus bounded how $X(T_n)$ departs from the identity, from which there remains to deduce a bound on the eigenvalues of
    $$\grad \flow_{T_n} = e^{- \kappa T_n} R_{-\delta T_n} X = e^{- \kappa T_n} R_{-\delta T_n} \;\; + \;\; e^{- \kappa T_n} R_{-\delta T_n}(X(T_n)-\mathbb{1}) \; .$$
    The last expression indicates how we intend to view $\grad \flow_{T_n}$, namely as the flow corresponding to $\beta=0$ plus a perturbation. The Bauer-Fike theorem (recalled as Theorem \ref{bauer_fike} below), bounds how eigenvalues behave under
    such perturbations. In this theorem, we use $p=\infty$ for the Schatten norm, which corresponds to the operator norm that we have used above. The norm of the perturbation in the right-hand side of \eqref{eq:BFcrit} is bounded by $\norm{e^{- \kappa T_n} R_{-\delta T_n} \qty(X(T_n)-1)} \leq e^{-\kappa T_n} \qty(e^{2 \betadiss T_n} - 1)$,
    and since $e^{- \kappa T_n} R_{-\delta T_n}$ is diagonalized by a unitary, the condition number equals 1. The Bauer-Fike theorem then says that for any eigenvalue $\mu$ of $\grad \flowdiss_{T_n}$, there exists an eigenvalue $\eta$ of $e^{- \kappa T_n} R_{-\delta T_n}$ such that
    \begin{equation}\label{eq:bound_bauer}
        \abs{\eta - \mu} \leq  e^{-\kappa T_n} \qty(e^{2 \betadiss T_n} - 1).
    \end{equation}
    Of course we know that $\eta = e^{(-\kappa \pm i \delta)T_n}$.

    The final argument is thus: if any eigenvalue $\mu$ is sufficiently close to some $\eta = e^{(-\kappa \pm i \delta)T_n}$, while all these $\eta$ are sufficiently far from $-1$, then each $\mu$ can be bounded away from $-1$. Explicitly, if
\begin{equation}\label{eq:AS0922crit}
    \abs{1 + e^{(-\kappa \pm i \delta)T_n}} > e^{- \kappa T_n} \qty(e^{2 \betadiss T_n} -1),
\end{equation}
    then
\begin{align}
        \abs{\mu + 1} &= \abs{\eta - e^{(-\kappa \pm i \delta)T_n} + e^{(-\kappa \pm i \delta)T_n} + 1}
                          \geq  \abs{1 + e^{(-\kappa \pm i \delta)T_n}} - \abs{\eta - e^{(-\kappa \pm i \delta)T_n}} > 0.
    \end{align}
Multiplying both sides of \eqref{eq:AS0922crit} by $e^{\kappa T_n}$, one readily sees that $\kappa=0$ is the most constraining case, and working out the algebra for this case gives the stated criterion \eqref{eq:ASNcrit1}.
\end{proof}

We here recall the two lemmas used in the proof of Lemma~\ref{lem:no_minusone_crit}:

\begin{lem}\label{gronwall_lemma}\emph{(Gr\"onwall,~\cite{book:98484})}
    Consider the integral equation
    \[h(t) \leq c(t) + \int_0^t g(s) h(s)\,\dd s,\]
    with the scalar functions $g,h$ and $c$ all non-negative on the interval $[0,t]$, and $c$ differentiable.
    Then
    \[h(t) \leq c(0) \exp(\int_0^t g(s) \, \dd s) + \int_0^t \dv{c}{t}\,\!(t) \qty(\exp(\int_s^t g(\tau) \, \dd \tau)) \, \dd s.\]
\end{lem}
We have used this Gr\"onwall Lemma with both $g$ and $c$ constant. In particular, the second term on the right drops out.

\begin{thm}\emph{(Bauer-Fike,~\cite{Bauer1960})}\label{bauer_fike}
    Suppose $A \in \mathbb{C}^{n \times n}$ is a diagonalizable matrix, and $V \in \mathbb{C}^{n \times n}$ is the non-singular similarity transformation that brings
    $A$ into its diagonal form $\Lambda$: \[\Lambda = V^{-1} A V.\]
    Define the condition number
    \[\kappa_p(V) := \frac{\norm{V}_p}{\norm{V^{-1}}_p},\]
    with $\norm{\cdot}_p$ the $p$-Schatten norm.
    Let $\mu$ be an eigenvalue of $A + B$, $B \in \mathbb{C}^{n \times n}$. Then there exists an eigenvalue $\eta$ of $A$ such that
    \begin{equation}\label{eq:BFcrit}
    \abs{\eta - \mu} \leq \kappa_p(V)\norm{B}_p.
    \end{equation}
\end{thm}
We used the Bauer-Fike Theorem with the operator norm ($p=\infty$) and with a matrix $A$ which is diagonalized by a unitary, such that $\kappa_p(V)=1$.\\

Lemma \ref{lem:no_minusone_crit} is useful only for part of the parameter space, represented by blue bands on Fig.~\ref{fig:nudregions}. Indeed, the criterion \eqref{eq:ASNcrit1} requires $\betadiss=0$ as $\bar\delta$ tends towards its maxial value $\frac{\nuddiss}{2n}$. Therefore, we next provide in Section \ref{sssec:globalnobeef} another result to cover these largest values of $\bar\delta$ (red bands on Fig.~\ref{fig:nudregions}). We will then combine the two in Section \ref{sssec:allnobeef} to obtain our overall conclusion.

\subsubsection{Values of $\betadiss$ and $\nuddiss$ excluding an $n$-subharmonic}\label{sssec:globalnobeef}

The rough idea can be sketched as follows.
We consider a harmonic $2 \pi / \nuddiss$-periodic solution as a point of reference $-$ we will show that such a solution must exist for any parameter values.
If the natural dynamics for $\betadiss = 0$ corresponds to a trajectory where a half-integer number of laps around the harmonic orbit are completed over a period $2n\pi/\nuddiss$,
then in case of small $\betadiss > 0$ it is unlikely for a trajectory which completes an integer number of laps around the harmonic solution to exist \emph{anywhere}.
It then immediately follows that it is unlikely for any trajectory to exist that can possibly close on itself to form a periodic orbit.
The only remaining trivial fixed point of $\pmap^n$ would be the unavoidable fixed point of $\pmap$ corresponding to the harmonic solution.

\begin{lem}\label{lem:nsubdisappears}
If the interval $[(1-\betadiss)\frac{1}{\nuddiss},\,(1+\betadiss)\frac{1}{\nuddiss}]$ contains no integer multiple of $1/n$, then the system \eqref{eq:vector_fieldh} can feature no $\frac{2 \pi n}{\nuddiss}$-periodic solution other than a single $2\pi/\nuddiss$-periodic solution.
\end{lem}
\begin{proof}
By Lemma\ref{lem:alwayslimcycle} in Section \ref{ssec:app:contractive}, the system always features at least one $2\pi/\nuddiss$-periodic solution. Let us denote it by ($x_1^*(s), p_1^*(s)$) and define the displaced variables
\begin{subequations}
\begin{align}
    x_d(s) = \xdiss(s) - x_1^*(s)\\
    p_d(s) = \pdiss(s) - p_1^*(s),
\end{align}
\end{subequations}
describing how other solutions behave compared to this trajectory. The corresponding (exact) equations of motion are
\begin{subequations}\label{eq:symmetric_dissipation}
\begin{align}
    \dv{\tdiss} x_d &= \hphantom{-} p_d - \kappa x_d,\\
    \dv{\tdiss} p_d &= - x_d - \kappa p_d - 2 \betadiss \sin(\frac{x_d}{2}) \cos(\frac{x_d}{2} + x_1^*(s) + \xi_d \sin(\nuddiss \tdiss)).
\end{align}
\end{subequations}
In polar coordinates $x_d = R \cos(\theta),\; p_d = R \sin(\theta)$, we obtain:
\begin{align}
    \dv{\tdiss} R &= -\kappa R - \sin(\theta)
    2 \betadiss \sin(\frac{R}{2}\cos(\theta)) \cos(\frac{R}{2}\cos(\theta) + x_1^*(s) + \xi_d \sin(\nuddiss \tdiss)) \; ,\\ \label{eq:ASlasttheta}
    \dv{\tdiss} \theta &= -1 - \betadiss \cos^2(\theta) \frac{ \sin(\frac{R}{2}\cos(\theta))}{\frac{R}{2}\cos(\theta)} \cos(\frac{R}{2}\cos(\theta) + x_1^*(s) + \xi_d \sin(\nuddiss \tdiss))\; .
\end{align}
By definition, $x_d(s) = p_d(s) = 0$ for any time $s$, corresponding to $R=0$, is a solution and no other trajectories ever cross the point $R=0$. Recognizing the expression $\frac{\sin(p_d/2)}{p_d/2} \in [-1,1]$ in \eqref{eq:ASlasttheta}, we can bound $\vert\dv{\tdiss} \theta+1 \vert \leq \betadiss$. A $\frac{2 \pi n}{\nuddiss}$-periodic solution has to make an integer number of laps $m$ around the (periodically displaced) origin, including possibly $m=0$. We thus need
$$ \theta(0)-\theta\qty(\frac{2 \pi n}{\nuddiss}) = 2 \pi m \in  \qty[\frac{2 \pi n}{\nuddiss} (1 - \betadiss) , \frac{2 \pi n}{\nuddiss} (1 + \betadiss)],$$
which is equivalent to the statement.
\end{proof}

Using the notation of Lemma \ref{lem:no_minusone_crit}, this criterion excludes the existence of an ($n$:$m$)-subharmonic 
\begin{equation}\label{eq:ASNcrit2}
    \text{ for  $|\delta^{(n:m)}| > \bar{\delta}$, provided } \frac{2 \pi n \betadiss}{\nuddiss} < \frac{2 \pi n \bar{\delta}}{\nuddiss}\;.
\end{equation}

\subsubsection{Combining Lemma \ref{lem:no_minusone_crit}  and Lemma \ref{lem:nsubdisappears} into a uniform criterion on $\beta$}\label{sssec:allnobeef}

We now select $\bar{\delta}$ to have overlapping regions of $1/\nuddiss$, as explained in Figure \ref{fig:nudregions}.

\begin{cor}\label{cor:final_no_minusone_crit}
Consider a simply connected open set $\mathcal{S}_p$ of possible parameter values $(1/\tilde{Q},\nu_d, \xi_{d})$ with $\nu_d \geq \nu_{\min}$ and $\tilde{Q}\geq Q_{\min} > 1/2$. For some parameter setting in this set, consider a stable subharmonic solution of~\eqref{eq:classical_system} of period smaller than $2\pi \bar n/\nu_d, \bar{n} \geq 2, \bar{n} \in \mathbb{N}$ and denote the largest period that this solution could possibly have when varying $\nu_d$ as $\bar\tau = 2\pi \bar n/\nu_{\min}$. If
\begin{equation}\label{eq:criterion}
\beta < \frac{0.53}{\bar\tau} \sqrt{1 - \frac{1}{4 Q_{\min}^2}},
\end{equation}
then this solution cannot undergo a period-doubling bifurcation when varying $(1/\tilde{Q},\nu_d,\xi_d)$ in $\mathcal{S}_p$.
\end{cor}
\begin{proof}
To cover the whole parameter space, we must match the value of $\bar{\delta}$ in both criteria, \eqref{eq:ASNcrit1} and \eqref{eq:ASNcrit2}. Selecting $\bar{\delta}$ smaller makes \eqref{eq:ASNcrit1} easier to achieve, yet it confines $1/\nuddiss$ close to integer multiples of $1/n$. Conversely, selecting $\bar{\delta}$ larger makes \eqref{eq:ASNcrit2} easier to achieve, while confining $1/\nuddiss$ closer to half-integer multiples of $1/n$. The best compromise is obtained when both criteria yield the same bound on $\betadiss$. We thus equate the right hand sides of \eqref{eq:ASNcrit1} and \eqref{eq:ASNcrit2}, to numerically find the optimal value $\bar{\delta} \simeq 0.537 \frac{\nuddiss}{2 \pi n}$ for the boundary between red and blue regions of Figure \ref{fig:nudregions}. Replacing this value in \eqref{eq:ASNcrit2}, working back to the original variables using~\eqref{eq:beta_scaling},~\eqref{eq:nu_d_scaling} and imposing the condition for all parameter values then gives the stated criterion.
\end{proof}

\noindent \textbf{Important remark:} Corollary \ref{cor:final_no_minusone_crit} speaks of a \emph{subharmonic} and this should be taken in the strict sense, i.e.~$n 2\pi/\nuddiss$-periodic with $n>1$. Indeed, when using Lemma \ref{lem:nsubdisappears} in the proof, we leave open the possible behavior of a $2\pi/\nuddiss$-periodic solution. This is however no problem for our intended use of Corollary \ref{cor:final_no_minusone_crit} to exclude a period-doubling cascade. Indeed, we thus leave open the possibility that a $2\pi/\nuddiss$-periodic solution would undergo period-doubling, but the resulting subharmonic would then be covered by Corollary \ref{cor:final_no_minusone_crit} such that further period-doubling is necessarily excluded.\\

Under the condition of Corollary \ref{cor:final_no_minusone_crit}, we are thus always in a zone where the considered subharmonic cannot undergo period-doubling, either thanks to Lemma \ref{lem:no_minusone_crit} or because it cannot exist in the first place (Lemma \ref{lem:nsubdisappears}). In other words, starting from a subharmonic of period $2\pi n/\nuddiss$ we are necessarily in the blue zone of Fig.~\ref{fig:nudregions}, and when moving towards the red zone of Fig.~\ref{fig:nudregions} this subharmonic must disappear without period-doubling.

Note that this result is independent of $\xi_d$ (essentially the drive amplitude) and only requires a practically trivial \emph{upper} bound on the dissipation $1/\tilde{Q}$.\\

Finally, to apply the Gambaudo-Tresser Conjecture \ref{conj:tresser}, the parameter set $\mathcal{S}_p$ of Corollary \ref{cor:final_no_minusone_crit}, with condition \eqref{eq:criterion}, should overlap with a zone where condition \eqref{eq:contrcondition} of Lemma \ref{lem:contraction} holds (green zones on Fig.~\ref{fig:opnochaos}). Indeed, this would put us in the conditions of Fig.~\ref{fig:opnochaos} (b): for some parameter region we know that there is a single harmonic limit cycle, while for an overlapping region we know that there can be no period-doubling cascade (starting below $\bar{\tau}$), and thus according to the conjecture there can be no transition into a chaotic regime.

Let us thus analyze how to combine both conditions \eqref{eq:contrcondition} and \eqref{eq:criterion}. Condition \eqref{eq:criterion} holds for all $Q_{\min} \geq \underline{Q}$ if it holds for $\underline{Q}$. Conversely, 
for $\tilde{Q}>1/\sqrt{2}$, the right hand side of \eqref{eq:contrcondition} is decreasing in $\tilde{Q}$, so \eqref{eq:contrcondition} holds for all $\tilde{Q} \leq \bar{Q}$ if it holds for $\bar{Q}$. For the regions satisfying \eqref{eq:contrcondition} and \eqref{eq:criterion} to overlap for a fixed $\beta$, we thus need $\underline{Q}<\bar{Q}$. By inspection, the limit is obtained at $1/\underline{Q} = 1/\bar{Q} = 0.53/\bar{\tau} < \sqrt{2}$. The associated constraint on $\beta$ becomes:
$$\beta < \frac{0.53}{\bar{\tau}} \, \sqrt{1-\left(\frac{0.53}{2\bar{\tau}}\right)^2} \; .$$\\

This concludes our results on avoiding a chaotic regime in the classical system \eqref{eq:classical_system}. Note that our arguments are mainly based on the fact that the Josephson potential and its first derivatives are uniformly bounded, so similar results should hold in other systems with these properties.

\section{Classical analysis of resonant behavior}\label{chap:resonance_analysis}

In the main text, we showed a clear correspondence between stable subharmonic solutions of period $\frac{2 \pi n}{\nu_d}$ and winding
number $m$ on the one hand, and the confinement of Schr\"odinger cat states with $n$ legs on the other hand.
Figures 3(a) and 4(a),(b) of the main text gave us a preliminary numerical account of how the drive parameters should be chosen
to render the multi-photon confinement process resonant, and regular, in the non-chaotic case of $\beta \simeq 0.5$.
In this section, we set out to obtain a simplified model that describes the resonance condition the drive parameters ($\nu_d, \xi_d$) need to satisfy
in order to create stable ($n:m$)-subharmonics. This analysis is performed using the method of geometric averaging.
We apply this perturbative technique in the non-chaotic case of $\beta \lesssim 0.5$, treating $\beta$ as a perturbation.
However, in contrast to Section~\ref{sec:no_period_doubling}, we will capture $\mathcal{O}\qty(\beta)$ effects, and only neglect effects of
$\mathcal{O}\qty(\beta^2)$ in first instance.

The section is outlined as follows.
After a short summary on the theory of geometric averaging for periodic systems, we obtain a class of averaged models that eliminates the dependence on
time, for any fixed ($n:m$), and discuss its symmetries.
Next, we numerically analyze the equilibrium point structure of this average model for the case of the ($3:1$)-resonance,
and compare to the numerical Floquet-Markov simulations.

\subsection{Summary of averaging theory}\label{sec:summary_averaging_theo}

We first summarize the method of geometric averaging (originally due to Krylov and Bogoliubov~\cite{KrylovNBogoliubov1935}). Consider a general $T$-periodic vector field describing the dynamics of a two-dimensional state variable $z \in \mathbb{R}^2$.
The oscillations of the vector field as a function of time $t$ are assumed to be \emph{fast} with respect to the magnitude of the vector field itself.
To capture this, we introduce a small positive dimensionless variable $\varepsilon \ll 1$,
and assume the following equations of motion for $z$,
\begin{equation}\label{eq:averaging_normal_form}
    \dot{z} = \varepsilon f(z,t) \qq*{,} z \in \mathbb{R}^2, t \in \mathbb{R},
\end{equation}
where $f$ is assumed of the same order as the frequency of its oscillations:
\[\norm{ T f} = \mathcal{O}(1).\]
Another way to interpret this condition, is that the state of the system should not change significantly on the timescale of the oscillations
present in the vector field.
To~\eqref{eq:averaging_normal_form} we can associate an \emph{autonomous averaged} vector field,
by neglecting its oscillations in time:
\[\bar{f}(\cdot) = \frac{1}{T} \int_0^T f(\cdot, t) \, \dd t.\]
Based on the average vector field, one defines the averaged system
\begin{equation}\label{eq:averaged_system}
    \dot{\bar{z}} = \varepsilon \bar{f}(\bar{z}) \qq*{,} \bar{z} \in \mathbb{R}^2,
\end{equation}
whose solutions $\bar{z}(t)$ are meant to approximate those of the true system~\eqref{eq:averaging_normal_form}, for small values of $\varepsilon$.
More precisely, the flow of the averaged system over one system-period provides a good approximation of the true Poincar\'e map
of~\eqref{eq:averaging_normal_form}.
We define
\begin{subequations}
    \begin{align}
        \pmap_\varepsilon & \coloneqq \flow_T \text{ with } \quad \pdv{t} \flow_t(z_0) = \varepsilon f({\flow}_t(z_0), t) \text{ and }\flow_0(z_0)=z_0,\quad \forall z_0 \in
        \mathbb{R}^2, t \in \mathbb{R}.
    \end{align}
\end{subequations}
A standard version of the averaging theorem for periodic systems can be found in chapter 4 of~\cite{guckenheimer2002nonlinear}.
We here summarize the main conclusions relevant to this work.
One can establish the following local correspondences between $\bar{f}$ and $\pmap_\varepsilon$, for small enough $\varepsilon$.

\begin{enumerate}[($i$)]
\item Consider a solution $\bar{z}(t)$ of~\eqref{eq:averaged_system} and a solution $z(t)$ of~\eqref{eq:averaging_normal_form}, based at $\bar{z}_0$ and $z_0$ respectively.
If $\abs{\bar{z}_0 - z_0} = \mathcal{O}(\varepsilon)$, then $\abs{\bar{z}(t) - z(t)} = \mathcal{O}(\varepsilon)$
on a timescale $0 < t < t_\textrm{max} = \mathcal{O}\qty(\frac{1}{\varepsilon})$.

\item Consider a \emph{hyperbolic equilibrium point} $\bar{z}^*$ of~\eqref{eq:averaged_system}, namely
\begin{equation}
    \bar{f}(\bar{z}^*) = 0,
\end{equation}
where the \emph{stability matrix} $A(\bar{z}^*) \coloneqq \grad\bar{f}(\bar{z}^*)$ only has eigenvalues with nonzero real parts.
Then there exists an $\varepsilon_0 > 0$ such that for all $0 \leq \varepsilon < \varepsilon_0$,
$\pmap_\varepsilon$ possesses a unique hyperbolic fixed point $z^*$ of the same stability type as $\bar{z}^*$, with $z^* = \bar{z}^* + \mathcal{O}\qty(\varepsilon)$.

\item Consider a trajectory $\bar{z}_s(t)$ in the stable manifold of the hyperbolic equilibrium point $\bar{z}^*$ of $\bar{f}$, and let $\pmap_\varepsilon^k(z_{s,0})$ be an
orbit in the stable manifold of the corresponding fixed point $z^*$ of $\pmap_\varepsilon$ (with still $z^* = \bar{z}^* +
\mathcal{O}\qty(\varepsilon)$).
If $\abs{\bar{z}_s(0) - z_{s,0}} = \mathcal{O}\qty(\varepsilon)$, then $\abs{\bar{z}_s(k T) - \pmap_\varepsilon^k(z_{s,0})} = \mathcal{O}\qty(\varepsilon),
\forall k \in \mathbb{N}$.
Similar results apply for unstable manifolds of hyperbolic fixed points (in reversed time).

\item If for some system parameter $\mu = \mu_0$, the averaged vector field $\bar{f}^{(\mu)}$ of~\eqref{eq:averaged_system} undergoes a saddle-node bifurcation,
then for $\varepsilon$ small enough, the Poincar\'e map $\pmap_\varepsilon^{(\mu)}$ of~\eqref{eq:averaging_normal_form} similarly undergoes a saddle-node bifurcation, for an $\varepsilon$-close parameter value $\mu_{0,\varepsilon}$.
\end{enumerate}

Since in the main text, the quantum-classical correspondence in the asymptotic regime was shown to involve stable periodic orbits of the Poincar\'e map,
we are interested in characterizing these as a function of system parameters.
The averaging theorem justifies characterizing the equilibrium points of an averaged vector field instead, if we can identify the corresponding small
parameter $\varepsilon$.
We will hence study the equilibrium points of the average vector field in a well-chosen frame.

\subsection{General properties of ($n:m$)-resonances}\label{sec:first_order_average}

We recall the form of the classical system with symmetric dissipation rates in the two quadratures, introduced in~\eqref{eq:symmetric_dissipation}:
\begin{subequations}\label{eq:symmetric_dissipation_recap}
\begin{align*}
    \dv{\tdiss} \xdiss &= \hphantom{-} \pdiss - \kappa \xdiss,\\
    \dv{\tdiss} \pdiss &= - \xdiss - \kappa \pdiss - \betadiss \sin(\xdiss + \xi_d \sin(\nuddiss \tdiss)).
\end{align*}
\end{subequations}
Here, the frequency of oscillation is given by the normalized drive frequency $\nuddiss$, and should be considered of order $1$.
To write our system in the normal form~\eqref{eq:averaging_normal_form}, amenable to averaging, we will move to a rotating frame with frequency
$\frac{m}{n} \nuddiss$, where $m$ and $n$ are two strictly positive coprime integers:
\begin{subequations}\label{eq:mn_rotframe}
\begin{align}
    \xdiss &= \cos(\frac{m}{n} \nuddiss \tdiss) u + \sin(\frac{m}{n} \nuddiss \tdiss) v,\\
    \pdiss &= \cos(\frac{m}{n} \nuddiss \tdiss) v - \sin(\frac{m}{n} \nuddiss \tdiss) u.
\end{align}
\end{subequations}
Recalling the definition of the detuning from~\eqref{eq:delta_def},
\begin{equation}\label{eq:delta_def_recap}
    \delta = 1 - \frac{m}{n} \nuddiss,
\end{equation}
the resulting equations of motion are
\begin{subequations}\label{eq:rotframe_carthesian}
\begin{align*}
    \dot{{u}} = \hphantom{-} \delta {v} - \kappa {u} + \betadiss { \sin(\frac{m}{n} \nuddiss \tdiss) \sin( {u} \cos(\frac{m}{n} \nuddiss \tdiss) + {v} \sin(\frac{m}{n} \nuddiss \tdiss) + \xi_d \sin(\nuddiss \tdiss))},\\
    \dot{{v}} = -\delta {u} - \kappa {v} - \betadiss { \cos(\frac{m}{n} \nuddiss \tdiss) \sin( {u} \cos(\frac{m}{n} \nuddiss \tdiss) + {v} \sin(\frac{m}{n} \nuddiss \tdiss) + \xi_d \sin(\nuddiss \tdiss))},
\end{align*}
\end{subequations}
where now $\delta$ can be considered small with respect to $\nuddiss$.
Note that we have changed the periodicity of the system, as the smallest common period of all time-dependent terms amounts to $2 n \pi / \nuddiss$.
The equilibria found through an averaged model will thus correspond to $2 n \pi / \nuddiss$-periodic subharmonic solutions a priori.
The averaged model is defined as
\begin{subequations}\label{eq:carthesian_averaged_model}
\begin{align}
    \dot{\bar{u}} = \hphantom{-} \delta \bar{v} - \kappa \bar{u} + \betadiss
    \frac{\nuddiss}{2 n \pi}\int_0^{\frac{2 n \pi}{\nuddiss}}  \sin(\frac{m}{n} \nuddiss \tdiss) \sin( \bar{u} \cos(\frac{m}{n} \nuddiss \tdiss) + \bar{v}
    \sin(\frac{m}{n} \nuddiss \tdiss) + \xi_d \sin(\nuddiss \tdiss)) \dd \tdiss,\\
    \dot{\bar{v}} = -\delta \bar{u} - \kappa \bar{v} - \betadiss
    \frac{\nuddiss}{2 n \pi}\int_0^{\frac{2 n \pi}{\nuddiss}} \cos(\frac{m}{n} \nuddiss \tdiss) \sin( \bar{u} \cos(\frac{m}{n} \nuddiss \tdiss) + \bar{v}
    \sin(\frac{m}{n} \nuddiss \tdiss) + \xi_d \sin(\nuddiss \tdiss)) \dd \tdiss.
\end{align}
\end{subequations}
For the solutions of this model to correspond to the true system up to good accuracy, we need to assume that
\begin{equation*}
    \kappa \sqrt{\bar{u}^2 + \bar{v}^2}       \ll \frac{\nuddiss}{n},\qquad
    \abs{\delta} \sqrt{\bar{u}^2 + \bar{v}^2} \ll \frac{\nuddiss}{n},\qquad
    \betadiss \ll \frac{\nuddiss}{n}.
\end{equation*}
The general correspondences between the averaged model and the true system summarized in the previous section are asymptotic in nature however,
so there are no clear a priori allowable values for ($\kappa, \delta, \beta$) (and corresponding regions in phase space) for which averaging is valid.
Explicit bounds on these values fall beyond the scope of this work. We will thus study the averaged model ~\eqref{eq:carthesian_averaged_model} \emph{as is},
knowing there exist some small-enough values for ($\kappa, \delta, \beta$) for which the  obtained
conclusions are valid for the true system.
In Section~\ref{sec:root_finding} however, we compare the predictions of this averaged model to exact numerical Floquet-Markov simulations.

The stability type of an equilibrium point ($u^*, v^*$) is defined in terms of the eigenvalues of the \emph{stability matrix }
\begin{align}
    A(\bar{u}^*, \bar{v}^*) &=
    \mqty(- \kappa  &  \delta   \\ - \delta  &  - \kappa)\label{eq:expression_stability_matrix} \\
    + \betadiss \frac{\nuddiss}{2 n \pi}
    &\mqty(\int_0^{\frac{2 n \pi}{\nuddiss}} \sin(\frac{m}{n} \nuddiss \tdiss) \cos(\frac{m}{n} \nuddiss \tdiss) \cos(\zeta(s)) \, \dd s
    & \int_0^{\frac{2 n \pi}{\nuddiss}} \sin^2(\frac{m}{n} \nuddiss \tdiss) \cos(\zeta(s)) \, \dd s \\
    - \int_0^{\frac{2 n \pi}{\nuddiss}} \cos^2(\frac{m}{n} \nuddiss \tdiss) \cos(\zeta(s)) \, \dd s
    & - \int_0^{\frac{2 n \pi}{\nuddiss}} \sin(\frac{m}{n} \nuddiss \tdiss) \cos(\frac{m}{n} \nuddiss \tdiss)\cos(\zeta(s)) \, \dd s),\nonumber
\end{align}
with \[\zeta(s) = \bar{u}^*\cos(\frac{m}{n} \nuddiss \tdiss) + \bar{v}^* \sin(\frac{m}{n} \nuddiss \tdiss)+ \xi_d \sin(\nuddiss \tdiss).\]
If both eigenvalues of $A$ have strictly negative real parts, ($\bar{u}^*, \bar{v}^*$) corresponds to a \emph{stable node}.
If $A$ has one eigenvalue with strictly positive real part and a second with strictly negative real part, we speak of a \emph{saddle point}.
If both eigenvalues of $A$ have strictly positive real parts, the equilibrium point corresponds to a \emph{source}.

The dissipative nature of~\eqref{eq:carthesian_averaged_model} dictates that 
equilibrium points ($\bar{u}^*,\bar{v}^*$) necessarily correspond to either stable nodes or saddle points, and no sources are allowed.
This is easy to see by considering $\Tr(A) \equiv - 2 \kappa$, so the eigenvalues of $A$ must sum to $- 2 \kappa$.
Furthermore, since $A$ only has real entries, its eigenvalues are either both real, or are a complex conjugate pair.
Then it easily follows that the eigenvalues $\eta_\pm$ of $A$ can be written as
\[\eta_\pm = - \kappa \pm \chi,\]
where $\chi$ is either strictly positive or purely imaginary.
Consequently, the only possible bifurcation mechanism is a \emph{saddle-node bifurcation}
where $\eta_+ = 0$, for $\chi = \kappa$.
At this bifurcation point, a \emph{saddle-node pair is either created or annihilated together} (depending in which direction one changes the system parameters).
Denoting the total number of nodes by $N_\textrm{n}$, and the total number of saddle points by $N_\textrm{s}$,
this implies that for any set of system parameters ($\kappa, \delta, \beta, \xi_d$), $N_\textrm{n} - N_\textrm{s}$  remains constant.
For system~\eqref{eq:carthesian_averaged_model}, we can prove that $N_\textrm{n} - N_\textrm{s} \equiv 1$.
The main idea behind the proof is a standard topological argument based on the Poincar\'e index (see Proposition 1.8.4 of~\cite{guckenheimer2002nonlinear})
of a closed curve $\mathcal{C}$ encircling all the equilibria\footnote{We show in Section~\ref{ssec:limiting_behavior_first_order_average} that for $\kappa > 0$ all equilibria are necessarily situated in a bounded region of phase space.}.
The index theorem equates $N_\textrm{n} - N_\textrm{s}$ to the number of turns made by the vector field when traversing the curve $\mathcal{C}$,
and for our system this number of turns amounts to $1$.



We now turn to finding the equilibrium points of~\eqref{eq:carthesian_averaged_model}.
It is instructive to perform the equilibrium point analysis in polar coordinates,
\begin{subequations}\label{eq:polar_coordinates}
\begin{align}
    \bar{u} & = R \sin(\theta),\\
    \bar{v} &= R \cos(\theta).
\end{align}
\end{subequations}
The equivalent vector field for ($\theta, R$) becomes
\begin{subequations}\label{eq:av_vf_polar}
\begin{align}
    \dot{\theta} &= \delta + \betadiss \frac{g\res(\theta, R, \xi_d)}{R}\label{eq:theta_dot_av},\\
    \dot{R} &= - \kappa R + \betadiss h\res(\theta, R, \xi_d)\label{eq:R_dot_av},
\end{align}
\end{subequations}
with
\begin{subequations}\label{eq:def_g_h}
\begin{align}
    g\res(\theta, R, \xi_d) &=
    \begin{cases}
    \hphantom{-} \sum_{k= - \infty}^\infty \cos(k n \theta) \mathcal{J}_{1 + k n}(R) \mathcal{J}_{- k m}(\xi_d) &\qq{,} m + n \textrm{ even},\\
    \hphantom{-} \sum_{k= - \infty}^\infty \cos(2 k n \theta) \mathcal{J}_{1 +2 k n}(R) \mathcal{J}_{- 2 k m}(\xi_d) &\qq{,} m + n \textrm{ odd},\\
    \end{cases}\\
    h\res(\theta, R, \xi_d) &=
    \begin{cases}
    - \sum_{k= - \infty}^\infty \sin(k n \theta) \mathcal{J}_{1 + k n}(R) \mathcal{J}_{- k m}(\xi_d) &\qq{,} m + n \textrm{ even},\\
    - \sum_{k= - \infty}^\infty \sin(2 k n \theta) \mathcal{J}_{1 + 2 k n}(R) \mathcal{J}_{-2 k m}(\xi_d) &\qq{,} m + n \textrm{ odd},
    \end{cases}
\end{align}
\end{subequations}
where $\mathcal{J}_l$ is the $l$'th order Bessel function of the first kind.
To make the notation more uniform, we will introduce the parity
\[r \coloneqq (m+n) \! \mod 2 = \begin{cases}
                               0 &\qq*{,} m + n \textrm{ even},\\
                               1 &\qq*{,} m + n \textrm{ odd},
                           \end{cases}
\]
and drop the superscript $\res$ to make the notation less heavy, while it should be remembered that $g$ and $h$ depend on the pair of coprime integers ($m,n$).
In this way we can write
\begin{subequations}\label{eq:new_gh_parity}
\begin{align}
    g(\theta, R, \xi_d) &= \hphantom{-} \sum_{k= - \infty}^\infty \cos(\pf k n \theta) \mathcal{J}_{1 + \pf k n}(R) \mathcal{J}_{- \pf k m}(\xi_d),\\
    h(\theta, R, \xi_d) &= - \sum_{k= - \infty}^\infty \sin(\pf k n \theta) \mathcal{J}_{1 + \pf k n}(R) \mathcal{J}_{- \pf k m}(\xi_d).
\end{align}
\end{subequations}
As an immediate observation,  for any value of $\kappa, \beta$ and $\delta$, if either $n \geq 2$, or $n = 1$ and $r = 1$, 
the origin $R = 0$ corresponds to an equilibrium point, since $h(\theta, 0 , \xi_d) = 0, \forall \theta, \xi_d$.
Since we are interested in finding subharmonic solutions with $n > 1$, we can exclude the case $n=1$, $r = 0$ however, so we can always
assume the origin to be an equilibrium point.
Subsequently excluding the origin, the remaining equilibria ($\theta^*, R^*$) can be sought for by solving
\begin{subequations}
\begin{align}
    \delta &= - \betadiss \frac{g(\theta^*, R^*, \xi_d)}{R^*},\label{eq:delta_forward}\\
    \kappa &= \hphantom{-} \betadiss \frac{h(\theta^*, R^*, \xi_d)}{R^*}\label{eq:kappa_forward}.
\end{align}
\end{subequations}
Thus excluding the origin, whenever we find a node (resp.\ saddle point) we know there must exist a corresponding saddle point (resp.\ node).
For this reason, we do not focus on the stability type for now, and postpone this question to Section~\ref{sec:root_finding}.
The rest of this section is outlined as follows.
First, a set of global symmetries of the set of equilibria is discussed.
Next, in Section~\ref{ssec:limiting_behavior_first_order_average} we consider some insightful limiting cases, for which analytical conclusions can be obtained.
Section~\ref{sec:root_finding} then details a numerical approach for characterizing the set of equilibria, for the case $m = 1, n = 3$.

\subsubsection{Global symmetries}\label{ssec:symmetries_first_order_average}

The averaged model~\eqref{eq:carthesian_averaged_model} adheres to a global rotational symmetry, causing a degeneracy in the set of equilibria.
\begin{enumerate}[($i$)]
    \item The averaged vector field is invariant under rotation by an angle $\frac{2 \pi}{ n \pf}$,
    since $h(\theta + \frac{2 \pi}{ n \pf}, \cdot, \cdot) = h(\theta, \cdot, \cdot)$, and
    $g(\theta + \frac{2 \pi}{ n \pf}, \cdot, \cdot) = g(\theta, \cdot, \cdot)$.
    This means that for any trajectory ($\theta(\tdiss), R(\tdiss)$), another solution is obtained by considering
    ($\theta(\tdiss) + k \frac{2 \pi}{ n \pf}, R(\tdiss)$), $k = 1, \ldots, \pf n - 1$.
    In particular, any equilibrium ($\theta^*, R^*$) is part of a group of $\pf n$ equilibria of \emph{the same stability type}.

    \item In the Hamiltonian limit of $\kappa = 0$, $\dot{R} = 0$ is automatically satisfied for
    \begin{equation}\label{eq:theta_angles_nodiss}
        \theta^* = \frac{k \pi}{ n \pf} , k = 0, \ldots, \pf n - 1,
    \end{equation}
    readily identifying a subset of possible equilibria. Note that for any value of $R^*$, the detuning can be chosen to
    satisfy~\eqref{eq:delta_forward}, so such equilibria must exist for
    \[
    \delta \in  \qty[- \betadiss g_\textrm{sup}, - \betadiss g_\textrm{inf}],
    \]
    where
    \begin{align*}
        g_\textrm{inf} &\coloneqq \inf_{\theta^* \in\qty{0, \frac{\pi}{n \pf}}, R > 0} \frac{g(\theta^*, R^*, \xi_d)}{R},\\
        g_\textrm{sup} &\coloneqq \sup_{\theta^* \in\qty{0, \frac{\pi}{n \pf}}, R > 0} \frac{g(\theta^*, R^*, \xi_d)}{R}
    \end{align*}
    Since $g$ is bounded, these limits are well-defined.

    \item Still in the Hamiltonian case, an extra symmetry of~\eqref{eq:av_vf_polar} can be established:
    \begin{subequations}\label{eq:time_reversal_symmetry}
    \begin{align}
        \theta & \rightarrow - \theta,\\
        s & \rightarrow - s.
    \end{align}
    \end{subequations}
    The system is therefore called \emph{reversible}, due to this time-reversal symmetry.
    The symmetry~\eqref{eq:time_reversal_symmetry} has no immediate extra consequences for any of the equilibria satisfying $\theta^* = k \frac{\pi}{n \pf}$,
    since the set is invariant under $\theta \rightarrow - \theta$.
    However, any equilibrium ($\theta^*, R^*$) that is not of this form must necessarily
    come with a second equilibrium ($-\theta^*, R^*$) of the same stability type.
    This symmetry is broken for $\kappa > 0$, but one can expect a certain approximate symmetry to hold, as an infinitesimal amount of dissipation can
    only change the phase portrait in a continuous manner.
\end{enumerate}


\subsubsection{Limiting behavior}\label{ssec:limiting_behavior_first_order_average}

There are two interesting limits to be considered in terms of the distance $R$ to the origin.
\begin{itemize}

\item For $R \ll 1$, we can Taylor expand~\eqref{eq:av_vf_polar} to obtain up to leading order, for $n > 1$:
\begin{subequations}
\begin{align}
    \dot{\theta} &= \delta + \frac{\betadiss}{2} \mathcal{J}_0(\xi_d) + \mathcal{O}\qty(R),\\
    \dot{R} &= - \kappa R + \mathcal{O}\qty(R^2).\label{eq:origin_is_stable}
\end{align}
\end{subequations}
The origin of phase space is seen to be a stable node for $\kappa > 0$, and a center in the limit of $\kappa \rightarrow 0$.
We call this equilibrium the \emph{nominal point}.
Close to the origin, the averaged model
describes an essentially linear system (harmonic oscillator). The Josephson nonlinearity only shows itself in the fact that
the frequency of this effective harmonic oscillator depends on the drive amplitude $\xi_d$, which is not the case for the response of a purely linear system.
We call this drive-induced frequency shift the \emph{AC-Stark shift} of the oscillator:
\begin{equation}\label{eq:ac_stark_shift}
    \Delta_\textrm{AC} \coloneqq \frac{\betadiss}{2} \qty(\mathcal{J}_0(\xi_d) - 1).
\end{equation}
One important conclusion is that the AC-Stark shift shows oscillatory behavior with $\xi_d$, and remains bounded  as a function of the drive amplitude $\xi_d$.

\item The opposite limit can also be taken. Consider a candidate equilibrium point ($\theta^*, R^*$) and let $R^* \rightarrow \infty$.
Since $g$ is a bounded function of $R$, we obtain
\[\dot{\theta} = \delta + \mathcal{O}\qty(\frac{1}{R^*}).\]
As $R^* \rightarrow \infty$, the value of $\abs{\delta}$ that is allowed for an equilibrium point to occur tends to zero.
In terms of the original drive frequency this implies
\[\nuddiss = \frac{n}{m} + \mathcal{O}\qty(\frac{1}{R^*}).\]
The effect of the nonlinearity of the Josephson junction thus effectively disappears, and we obtain a resonance condition based solely on the
linear part of the system.
We can apply the same reasoning to the dissipation rate $\kappa$, as also $h$ is bounded in $R$.
For an equilibrium point to occur at a distance $R^*$, we need
\[\kappa =  \mathcal{O}\qty(\frac{1}{R^*}).\]

\item
There is one more limit that allows for a simple analytical estimate, namely the limit of weak driving: $\xi_d \ll 1$.
In this case we can replace the Bessel functions by an appropriate asymptotic expansion to obtain the leading-order contributions in~\eqref{eq:av_vf_polar}.
The asymptotic expansion
\begin{equation}\label{eq:asymptotic_expansion_bessel}
    \mathcal{J}_l(\xi_d) \sim \frac{\xi_d^l}{2^l l!} \qq*{,} l,\xi_d > 0
\end{equation}
is valid for $\xi_d \rightarrow 0$.
We obtain the following leading-order equation:
\begin{subequations}
\begin{align}
    \dot{\theta} &= \delta + \betadiss \frac{\mathcal{J}_1(R)}{R} + \mathcal{O}\qty(\frac{\xi_d^{\pf m}}{2^{\pf m} (\pf m)!}),\\
    \dot{R} &= - \kappa R - {(-1)}^{m} \betadiss \sin(\pf n \theta) \frac{\xi_d^{\pf m}}{2^{\pf m} (\pf m)!} \qty(\mathcal{J}_{\pf  n+1}(R) + {(-1)}^{r} \mathcal{J}_{\pf  n-1}(R))\nonumber\\
    &+ \mathcal{O}\qty(\frac{\xi_d^{2 \pf  m}}{2^{2 \pf  m} (2 \pf  m)!}).\label{eq:rdot_low_xi}
\end{align}
\end{subequations}
In this case, it is $\delta$ that first determines $R^*$ such that $- \delta / \betadiss \simeq \mathcal{J}_1(R^*) / R^*$ (though possibly not uniquely),
and the value of $\theta$ can then be chosen as to satisfy $\dot{R} = 0$ in~\eqref{eq:rdot_low_xi}.
Note that this last equation always allows for a solution $\theta^*$ as long as the dissipation rate $\kappa$ is small enough.
In the limit of $\kappa = 0$, $\theta^* = \frac{k \pi}{n \pf}, k = 0, \ldots, \pf n-1$ provides a class of solutions.
However, for decreasing $\xi_d$, the averaged model tells us that the oscillator must be increasingly high-Q to observe the corresponding subharmonic resonance.
Indeed, using the boundedness of $\mathcal{J}_{\pf n \pm 1}(R) / R$ for $n > 1$, it easily follows from~\eqref{eq:rdot_low_xi} that
\begin{equation*}
    \kappa < \frac{\betadiss \xi_d^{\pf m}}{2^{\pf m-1} (\pf m)!} +  \mathcal{O}\qty(\frac{\xi_d^{2 \pf  m}}{2^{2 \pf  m} (2 \pf  m)!})
\end{equation*}
must be satisfied for there to exist any equilibria.
Note that odd-parity processes ($\parity = 1$) are suppressed, in the sense that asymptotically, for small $\xi_d$, one needs a much higher-Q system to observe them.
\end{itemize}


The general equilibrium point structure of the averaged model is quite complicated, and especially the sequence of bifurcations upon varying parameters is
increasingly complicated. In the next subsection, we adopt a numerical approach to characterize this equilibrium point structure as a function of the drive parameters.

\subsection{Numerical approach for ($n:m$) $=$ (3:1)}\label{sec:root_finding}

Note that for any fixed equilibrium ($\theta^*, R^*$), and any given drive amplitude $\xi_d$,
one obtains the corresponding values of $(\kappa, \delta)$ by \emph{forwardly} computing them from~\eqref{eq:av_vf_polar}.
If on the other hand, we want to fix the system parameters beforehand, and look for equilibrium points,
one has to resort to a numerical root-finding algorithm.
To avoid having to use root-finding algorithms in the two variables ($\theta, R$) we make an observation.
Note that the dissipation rate $\kappa$ is fixed at the fabrication of the device, whereas $\delta$
corresponds to a drive-detuning that is typically adjusted online.
This allows us to adopt the following strategy to find the equilibria
of~\eqref{eq:av_vf_polar} as a function of ($\betadiss, \kappa, \delta, \xi_d$), using a reliable \emph{one-dimensional} root-finding algorithm.
First fix the values of $\betadiss, \kappa$ \emph{and} the $R^*$-value of the sought-for equilibrium.
Next, from~\eqref{eq:R_dot_av}, root-find possible $\theta^*$-values that give rise to $\dot{R} = 0$.
For this we used a simple algorithm based on sign changes of $\dot{R}$ as function of $\theta$.
If no such sign changes are found,
we conclude that no equilibria exist for the given values of ($\kappa,\delta, R^*$).
For every $\theta^*$ that does give rise to $\dot{R} = 0$, from~\eqref{eq:theta_dot_av}, we can forwardly compute the (unique) value
\[\delta = - \betadiss \frac{g(\theta^*, R^*, \xi_d)}{R^*}\]
such that also $\dot{\theta} = 0$ and ($\theta^*, R^*$) is indeed an
equilibrium point.
As a note aside, the point $R^* = 0$ should be omitted, as $\theta$ is ill-defined at the origin.
From~\eqref{eq:origin_is_stable}, we know that the origin is always an equilibrium point however, and corresponds to a stable node when $\kappa > 0$.

Afterwards, the stability type of the corresponding equilibrium can be determined
by performing a linearization analysis on~\eqref{eq:carthesian_averaged_model}.
We recall that if the eigenvalues of $ A(R^* \sin(\theta^*) , R^* \cos(\theta^*)) $ (as defined in~\eqref{eq:expression_stability_matrix})
both have strictly negative real parts, then ($\theta^*, R^*$) corresponds to a stable node. 
If one of the eigenvalues has a strictly positive real part, then ($\theta^*, R^*$) corresponds to a saddle point.
In practice, it suffices to compute the determinant of~\eqref{eq:expression_stability_matrix}.
This determinant is  evaluated numerically in $(\theta^*, R^*)$.


\begin{figure}[H]
\begin{center}
\includegraphics[width=0.60\textwidth]{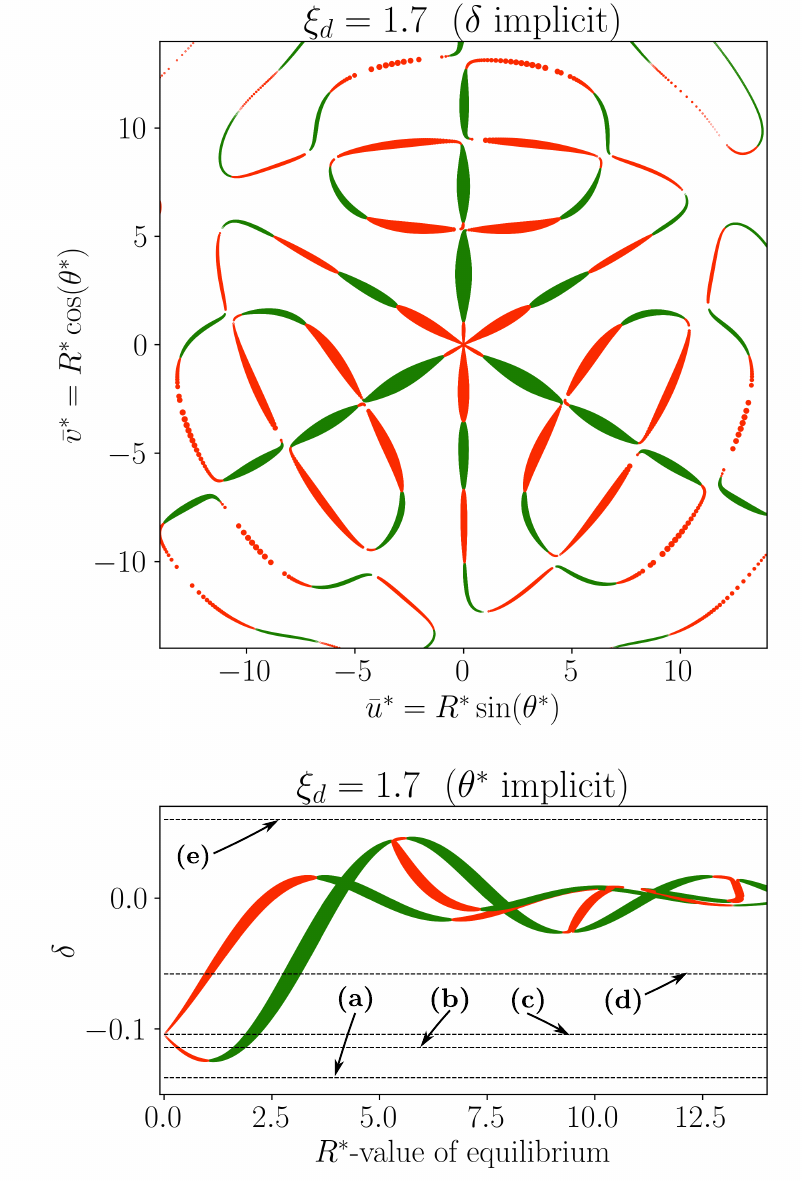}
\caption{Numerical account of the bifurcation structure of the averaged model~\eqref{eq:av_vf_polar},\eqref{eq:new_gh_parity}
when varying the detuning $\delta$, for $\betadiss = 0.5$, $\kappa = 10^{-5}$, and $\xi_d = 1.7$.
The possible equilibria ($\theta^*, R^*$) were determined by imposing $\dot{R} = 0$ (see~\eqref{eq:kappa_forward}).
Two projections are given for the remaining relation $\delta = - \betadiss g(\theta^*, R^*, \xi_d) / R^*$, relating the detuning $\delta$ to the polar coordinates ($\theta^*, R^*$) of the equilibrium point.
\textbf{(top):} The possible equilibria satisfying $\dot{R} = 0$ (see~\eqref{eq:kappa_forward}) are  plotted, omitting the corresponding (uniquely defined) value of $\delta$.
Stable nodes are shown in green, while saddle points are shown in red.
One can clearly see the rotational symmetry over an angle $\frac{2 \pi}{n (1 + r)} = 2 \pi / 3$ of the equilibrium point structure, which was introduced in Section~\ref{ssec:symmetries_first_order_average}.
Equilibria close to the origin ($R < 5$ here) occur approximately at angles $\theta = k \pi / 3, k = 0, \ldots 5$.
\textbf{(bottom):} $\delta$ is plotted as a function of $R^*$, omitting the corresponding value of $\theta^*$. 
Correspondingly, every displayed point corresponds to a triplet of equilibria, at the same distance $R$ from the origin, and angles $\pm 2 \pi / 3$ apart.
At local extremal points of $\delta$, a saddle-node bifurcation takes place, since three nodes and three saddles locally (dis)appear at these extrema.
The bifurcation mechanisms from situation (a) to (e) are discussed in the text.
\label{fig:saddle_nodes}}
\end{center}
\end{figure}

\subsubsection{Bifurcation structure}\label{ssec:bifurcation_structure}

In Figure~\ref{fig:saddle_nodes}, an account is given of the equilibrium point structure of the averaged model for $m = 1, n = 3$ (so an even-parity process, $\parity = 0$),
as a function of the detuning $\delta$, for fixed values of $\kappa = 10^{-5}, \betadiss = 0.5$ and $\xi_d = 1.7$.
Two projections of the relation 
\[\delta = - \betadiss \frac{g(\theta^*, R^*, \xi_d)}{R^*}\]
are given, linking the triplet ($\delta, \theta^*, R^*$).
Stable nodes are shown in green, while saddle points are shown in red.
On the top plot, the value of $\delta$ is implicit.
One can clearly see the rotational symmetry over an angle $\frac{2 \pi}{n (1 + r)} = 2 \pi / 3$ of the equilibrium point structure, which was introduced in Section~\ref{ssec:symmetries_first_order_average}.
On the bottom plot, the value of the angle $\theta^*$ is implicit.
At local extrema of $\delta$, a saddle-node bifurcation takes place, where three nodes and three saddles locally (dis)appear at these extrema.
Combining the bottom and top plots of Figure~\ref{fig:saddle_nodes},
we can discuss the location of the equilibria and the bifurcations that take place, when increasing $\delta$ gradually from point (a) to point (e).
\begin{itemize}
\item For the most negative values of $\delta$, no equilibria are found for $R > 0$. The only equilibrium resides in the origin (not shown), and the averaged model
      predicts the origin to be globally attractive for any $\kappa > 0$. This is straightforward to prove due to the absence of other equilibria.
\item Upon increasing the value of $\delta$, from point (a) to point (b), a threefold saddle-node bifurcation (around $R^* \simeq 1$) takes place,
      simultaneously creating three saddle-node pairs at angles $\theta^* \simeq 2 k \pi / 3, k = 0,1,2$.
      Upon further increasing the value of $\delta$, the stable nodes move radially outward to larger values of $R^*$, while the saddle points move radially inward towards the origin.
\item From point (b) to (d), the saddle points move through the origin, re-exiting at angles $\theta^* \simeq \pi / 3 + 2 k \pi / 3, k = 0,1,2$
      (note that in the origin $\theta^*$ can undergo a discontinuity).
      In situation (c), the three saddle points coalesce in the origin, creating a degenerate point
            \footnote{One can show that for $\kappa = 0$ this indeed corresponds to a degenerate bifurcation point where the $4$ equilibria exactly coincide,
            due to the global rotational symmetry of the vector field.
            For small non-zero $\kappa$ this bifurcation must be regularized in saddle-node bifurcations very close to this degenerate point,
            providing an \emph{unfolding} of this degenerate point in codimension $2$, similar in nature to the cusp catastrophe (see e.g.\ Section 7.1 of~\cite{guckenheimer2002nonlinear}).
            The exact nature of the unfolding of this bifurcation in terms of the two parameters ($\kappa, \delta$) has not been the study of this work however.}.
      The value of $\delta$ for situation (c) exactly compensates the AC-Stark shift (see~\eqref{eq:ac_stark_shift}) of the oscillator:
      \[\delta = - \frac{\betadiss}{2} \mathcal{J}_0(\xi_d) \simeq -0.1.\]
\item When further increasing $\delta$, moving from situation (d) to situation (e), many different saddle-node pairs are created and move towards other equilibria as a function of the detuning $\delta$,
      before subsequently recombining and annihilating in another saddle-node bifurcation.
      Moreover, we see that equilibria are no longer restricted to the angles $\theta \simeq\frac{k \pi}{3}, k = 0,\ldots 5$,
      as equilibria branch off and move in the angular direction in between these radial axes. Beyond the numerical results displayed in Figure~\ref{fig:saddle_nodes},
      we have no further analytical insight in this complicated bifurcation structure.
      This process continues when further increasing $\delta$, until eventually all equilibria are annihilated, corresponding to situation (e).
\end{itemize}
A schematic representation of the phase diagrams corresponding to the analogous situations (a) to (e) for the dissipationless case of $\kappa = 0$ is given in
Figure~\ref{fig:phase_portraits} (top).

The bifurcation structure for different values of $\xi_d$ looks qualitatively similar.
Specifically, we observe that for values of $\xi_d \in \qty[0,2.2]$, when driving to compensate the AC-Stark shifted frequency of the oscillator ($\delta = - \betadiss \mathcal{J}_0(\xi_d) / 2$)
there exists a \emph{unique} triplet of stable nodes (not shown here).
The distance $R^*$ of these nodes to the origin moreover increases monotonically from $0$ to $3.3$ with $\xi_d$ going from $0$ to $2.2$,
and for the corresponding angles we have $\theta^* \simeq 2 k \pi / 3, k = 0,1,2$.
This is true independently of the value of $\betadiss$, for small enough dissipation rate $\kappa$.
Driving the oscillator around its AC-Stark shifted frequency thus identifies one favorable working regime in which the global symmetries of~\eqref{eq:theta_angles_nodiss} approximately hold,
and in which the stable resonant nodes are unique.
In the next section, we will indeed see that the drive parameters giving rise to
a confined manifold of Schr\"odinger cat states for the quantum system generally follow the AC-Stark shift:
\[\delta = - \frac{\betadiss}{2} \mathcal{J}_0(\xi_d).\]

\subsubsection{Comparison to Floquet-Markov simulations}\label{ssec:comparison_floquet_markov}

In this section, we establish how the equilibrium point structure of the averaged classical model~\eqref{eq:av_vf_polar},\eqref{eq:new_gh_parity}
can predict the asymptotic behavior of the quantum system,
as calculated by the numerical Floquet-Markov simulations outlined in Section~\ref{sec:floquet_new_label}.
For this we consider the dissipationless case of $\kappa = 0$,
as the Floquet-Markov simulations are set in the limit of a vanishing coupling rate to a thermal bath, and hence at a vanishing dissipation rate.
The value of $\beta$ is chosen as to be in a regular regime, $\beta = 0.5$.
We again consider the case of the ($3:1$)-resonance as a guiding example, while the discussion can readily be applied to other resonances.
\begin{figure}[H]
\begin{center}
\includegraphics[width=0.7\textwidth]{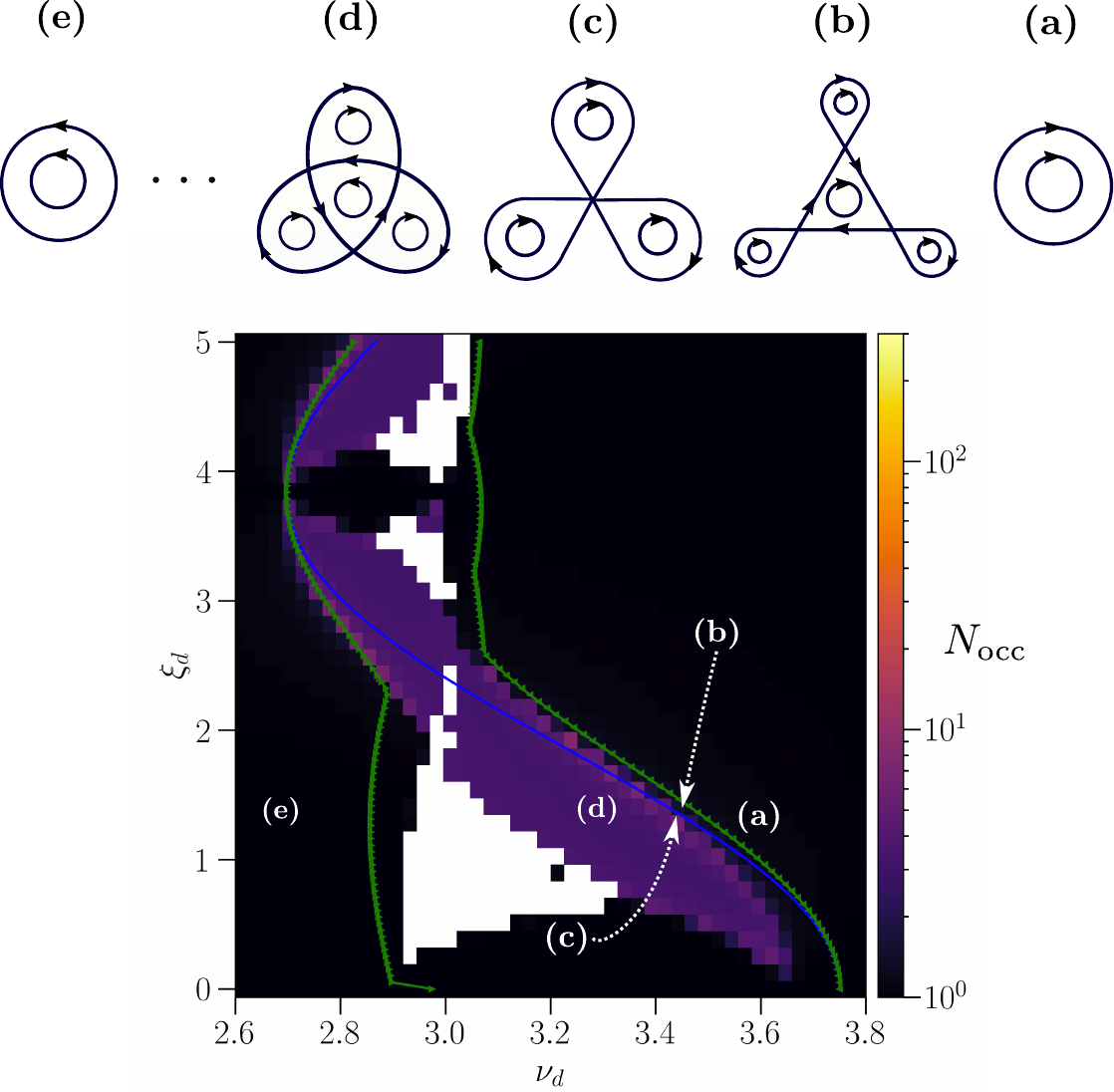}
\caption{Comparison between the Floquet-Markov signatures of the ($3:1$)-resonant cat states, and the occurrence of ($3:1$)-fixed points of
$\pmap^3$, as predicted by the averaged classical system~\eqref{eq:av_vf_polar},
for $\beta = 0.5$, $\lambda = 0.2$, and $\kappa = 0$.
\textbf{(Bottom):} The effective number of Floquet modes $N_\textrm{occ}$ occupied by $\rho_\infty$ at time $\tau = 0 \mod 2 \pi / \nu_d$ is plotted in a color map plot,
as a function of drive parameters (exact numerical Floquet-Markov simulations).
The green lines delimit the region where the classical averaged model~\eqref{eq:carthesian_averaged_model} predicts the existence of stable equilibria,
and is seen to delimit the quantum resonance region where $N_\textrm{occ} \simeq 3$ up to very good accuracy. The AC-Stark-shifted drive frequency as predicted by the averaged model (see~\eqref{eq:31_ac_stark_shift}) is shown in blue.
\textbf{(Top):} Schematic representation of the phase portraits of the averaged system for 5 points indicated in the bottom plot.
The center in the middle corresponds to the origin of phase space.
Moving from (a) to (e), the following bifurcation mechanisms take place.
First, a saddle-node bifurcation occurs where three nodes and three saddles are created, represented in (b). The saddle-separatrices form (initially small) homoclinic loops.
The corresponding saddle-points subsequently move radially inwards (typically very fast as a function of $\nu_d$), until they coalesce (and perfectly coincide, due to the discrete rotational
symmetry) in the origin. This is depicted in the degenerate situation (c), corresponding to the drive frequency $\nu_d$ that compensates for the AC-Stark shift, indicated by the blue line.
At this point, the direction of rotation of the origin is reversed, and in (d) the three saddle-points form heteroclinic loops. The bifurcation process from (d) to (e) involves (often cascades of) saddle-node annihilations and creations (see Figure~\ref{fig:saddle_nodes}),
when further lowering the value of $\nu_d$ (corresponding to increasing $\delta$, see~\eqref{eq:delta_def_recap}).
This entails exchanges in pairing between saddle-node triplets, before they eventually all cancel out, and only the center in the origin remains.
This situation is depicted in situation (e).
\label{fig:phase_portraits}}
\end{center}
\end{figure}
The general correspondence that can be established between the classical and quantum system was outlined in the main text.
For the case of the ($3:1$)-resonance, we saw that a stable $3$-orbit
\[\qty{\qty(x^{(3:1)}_l,p^{(3:1)}_l) \vert l = 0,1,2} \]
of the Poincar\'e map $\pmap$ corresponds to a triplet of Floquet modes
that are of the form of three-component Schr\"odinger cat states, and whose quasienergies were shown to be degenerate modulo $\nu_d / 3$,
indicating a multi-photon process where three oscillator photons are converted into one drive photon and vice versa.
We recall that these Schr\"odinger cat states are given by the superpositions of distinguishable states in phase space that closely resemble coherent states
$\ket{\alpha_l}$, $l = 0,1,2$.
The amplitudes $\alpha_l$ of these coherent states were seen to approximately correspond (see e.g.\ Figure 3 (c) and (e) of the main text) to the three classical fixed points of $\pmap^3$,
\begin{equation}\label{eq:alpha_correspondence}
    \alpha_l \simeq \frac{x^{(3:1)}_l + i p^{(3:1)}_l}{\sqrt{2}}\qq*{,} l = 0,1,2.
\end{equation}
In this section, we have described a prominent class of stable $(3:1)$-subharmonics corresponding to stable nodes
\[(R^* \sin(\theta^* + 2 l \pi / 3), R^* \cos(\theta^* + 2 l \pi / 3)), l = 0,1,2.\]
of the first-order averaged model~\eqref{eq:carthesian_averaged_model},
which approximate solutions of the true system.
Working back to the lab frame, by undoing the rotating-frame transformation~\eqref{eq:mn_rotframe},
we obtain that the corresponding $3$-orbit of the Poincar\'e map $\pmap$ corresponding to~\eqref{eq:symmetric_dissipation_recap}
is approximately of the form  
\begin{subequations}\label{eq:lab_frame_averaged_subharmonic}
\begin{align}
    \xdiss^{(3:1)}_l \simeq R^* \sin(\theta^* + 2 l \pi / 3),\\
    \pdiss^{(3:1)}_l \simeq R^* \cos(\theta^* + 2 l \pi / 3),
\end{align}
\end{subequations}
Recall from Section~\ref{sec:normal_form} that both quadratures $\xdiss,\pdiss$ were scaled by the
quantum scaling parameter (see ~\eqref{eq:AS:scaling}) $\pzpf$ to eliminate it from the classical equations of motion.
The correspondence~\eqref{eq:alpha_correspondence} can then be worked out for the $3$-orbit of the form~\eqref{eq:lab_frame_averaged_subharmonic},
yielding
\begin{equation}\label{eq:alpha_correspondence_R}
    \alpha_l \simeq i e^{- i \qty(\frac{2 \pi l}{3} + \theta^*)}\frac{R^*}{2 \pzpf}, l = 0,1,2.
\end{equation}

With this correspondence in mind, we can now compare the classical averaged model to the numerical Floquet-Markov simulations of the infinite-time behavior of the quantum system.
This is the subject of Figure~\ref{fig:phase_portraits} (bottom).
To characterize the asymptotic regime,
as in the main text, we consider the effective number of Floquet modes $N_\textrm{occ}$ occupied
by the asymptotic state $\rho_\infty(\tau)$, defined as 
\begin{equation*}
N_\textrm{occ} \coloneqq \exp(S(\rho_\infty)),
\end{equation*}
where $S(\rho_\infty)$ is the von Neumann entropy of the state,
\[S(\rho_\infty) \coloneqq - \Tr(\rho_\infty \ln(\rho_\infty)).\] 
$N_\textrm{occ}$ is plotted as a function of the drive parameters ($\nu_d, \xi_d$) in Figure~\ref{fig:phase_portraits} (bottom).
From a numerical point of view, we define the \emph{resonance region} for the quantum system to be the set of drive parameters ($\nu_d, \xi_d$) for which
\[N_\textrm{occ} \simeq 3,\]
since when $\rho_\infty$ is given by a uniform mixture of 3 three-component Schr\"odinger cat states,
$N_\textrm{occ}$ amounts to exactly $3$.
This quantum resonance region corresponds to the purple region in Figure~\ref{fig:phase_portraits}.
The black region corresponds to an essentially pure state (displaced vacuum) for $\rho_\infty$.
For the drive parameters corresponding to the white points, the numerical simulations are inconclusive to determine $\rho_\infty$ (see Remark~\ref{rem:non_convergence_new}).

Superimposed on the plot is the
blue line, representing the drive parameters that exactly compensate for the AC-Stark shift of the oscillator, given by
\begin{equation}\label{eq:31_ac_stark_shift}
    \nu_d^{(3:1)} \coloneqq \frac{3}{1} \qty(1 + \frac{\beta}{2} \mathcal{J}_0(\xi_d)).
\end{equation}
We can see that the general shape of the resonance region follows the oscillation of this AC-Stark shifted resonance frequency.
Plotted in green are the extremal values for the drive-detuning for which the averaged model exhibits stable nodes.
These necessarily coincide with a saddle-node bifurcation of the average system,
and analogously to the \emph{global} extrema of the analogous curves shown in Figure~\ref{fig:saddle_nodes} (bottom).
We can see that the green lines delimit the quantum resonance region ($N_\textrm{occ} \simeq 3$) up to very good approximation,
and hence the averaged classical model accurately predicts the drive parameters that lead to a resonant situation for the quantum system.
As a point of reference, the top of Figure~\ref{fig:phase_portraits} shows schematic diagrams of the corresponding phase-portrait of the classical averaged model (for $\kappa = 0$).
Examples of drive parameters that give rise to the respective phase portraits are indicated on the bottom plot of Figure~\ref{fig:phase_portraits}.
The drive parameters are chosen such that the averaged model predicts either a unique triplets of resonant equilibria (situations (b),(c),(d)),
or that only the origin remains as a center (situations (a) and (e)).

In conclusion, we can see that
the general shape of the quantum resonance region in terms of drive parameters is well-described by the classical averaged model of~\eqref{eq:carthesian_averaged_model}.
Since~\eqref{eq:carthesian_averaged_model} corresponds to the lowest-order model of the perturbative method of averaging,
and the value of $\beta$ chosen is quite large to be considered a perturbation
(but small enough to be in the regular, non-chaotic regime, $\beta = 0.5$) was not a priori given.
A first limitation of any classical model in predicting the quantum resonance region,
is that the classical equations of motion~\eqref{eq:symmetric_dissipation_recap} do not show any dependence on the quantum scaling parameter $\pzpf$,
which has a clear quantum effect studied in Figure 4 of the main text. We conclude with a few remarks on other limitations of classical models in predicting the resonance region for the quantum system,
leading to some expected quantitative differences between the purple region and the green delimiting lines in Figure~\ref{fig:phase_portraits}.

\begin{rem}\emph{(Tunneling and far equilibria)}. 
    A first aspect to be noted, is that the maximal number of Floquet modes occupied by $\rho_\infty$ is seen to be $N_\textrm{occ,max} \simeq 3$.
    The possible Floquet modes that are occupied correspond to either three $3$-component cat states (purple region, $N_\textrm{occ} \simeq 3$),
    or a single Floquet mode resembling a (dressed) vacuum state (black region, $N_\textrm{occ} \simeq 1$)
    \footnote{The white points in Figure~\ref{fig:phase_portraits}(bottom) indicate values for the drive parameters for which the numerical Floquet Markov simulations are inclusive to determine $\rho_\infty$,
    since for these drive parameters, both a complete mixture of the three cat states on the one hand, and the dressed vacuum state on the other hand appear to be long-lived,
    and both span the kernel of the transition matrix $R$ (see~\eqref{eq:rate_matrix}) obtained from Fermi's golden rule outlined in Section~\ref{sec:floquet_new_label}.
    For a more detailed discussion, see Remark~\ref{rem:non_convergence_new}.}.
    Note that this is not what would be predicted by the classical averaged model in the equivalent limit of vanishing dissipation rate,
    since many resonant stable nodes can co-exist (see Figure~\ref{fig:saddle_nodes}(bottom)), and all have a non-vanishing basin of attraction.
    The reason for this limitation of the quantum-classical correspondence in the asymptotic regime
    is the specific dissipation model for the driven quantum system, consisting of a weak Hamiltonian coupling to a thermal bath.
    Indeed, by the extension of Fermi's golden rule to periodically-driven systems outlined in Section~\ref{sec:floquet_new_label}, 
    the driven quantum system can directly transition between different Floquet modes, and is seen to converge to a mixture of Floquet modes that correspond
    to only a handful of the possible classical locally-stable nodes. 
    When comparing the phase-space representation of $\rho_\infty$ to the phase portrait of the classical Poincar\'e map, as in Figure 3 (c),(e) of the main text,
    we observe (simulations not shown here) that the dominantly-occupied Floquet modes in the resonant case always corresponds
    to the classical stable resonant nodes that are \emph{closest to the origin}, so in terms of the averaged model, have minimal values of $R^*$. This observation is in accordance with the energetic expectations as we assume a coupling to a zero temperature thermal bath. 
    This fact should also find a quantitative explanation through the application of semiclassical quantization methods based on the effective potential landscape associated
    to the nested irrational tori of the Poincar\'e map~\cite{Bensch1992,Breuer1991}.
    For the regular, non-chaotic regime of periodically driven systems in a Floquet-Markov-limit similar to this work,
    a Boltzmann distribution with a winding-number-dependent effective temperature has been shown to predict the distribution of Floquet modes~\cite{Ketzmerick2010}.
    Such an analysis falls beyond the scope of this letter however.
    As a last point, note that far-away resonant stable nodes might become relevant for the \emph{transient}, finite-time behavior of the system. This is not the subject of this work however.
\end{rem}

\begin{rem}\emph{(Basins of attraction)}. 
    A second main difference between the classical averaged model and the behavior of the quantum system lies in the fact that quantum states occupy a finite minimal surface area in phase space
    given by the Heisenberg uncertainty principle.
    In this way, classical stable nodes with a small basin of attraction~\footnote{
        While the classical averaged model was analyzed for $\kappa = 0$, we refer to the resulting centers as ``nodes'' after all,
        since for any $\kappa >0$, these correspond to stable nodes. Similarly, we refer to regions in phase space delimited by saddle separatrices as ``basins of attraction'' of the nodes,
        since in the limit of $\kappa \rightarrow 0$ but $\kappa > 0$, the true basins of attraction of the corresponding stable nodes are delimited by these saddle-separatrices.
    }  cannot be resolved by the quantum system, as the quantum wave-packet is forced to spread outside of the basin of attraction.
    Concretely, if the basin of attraction of a stable node is smaller than the surface in phase space occupied by a quantum coherent state $\ket{\alpha_l}$,
    the correspondence in the main text should be expected to break down.
    Such small basins of attraction for the classical system occur close to
    a saddle-node bifurcation point for example, and we cannot expect the corresponding resonant stable nodes to give rise to a resonant quantum regime,
    The same argument holds when the resonant nodes are sufficiently close to the origin, as the quantum state spreads over multiple classical basins of attraction (not shown here).
    This potentially explains the black region around $\xi_d \in \qty[0, 0.3], \nu_d \in \qty[3.67, 3.73]$,
    as $\nu_d$ is close to the saddle-node bifurcations of the green curves, and moreover, for small values of $\xi_d$, the stable nodes are created close to the origin (not shown here).
\end{rem}

\begin{rem}\emph{(Hilbert space truncation)}.
    Lastly, from~\eqref{eq:alpha_correspondence_R}, we can see that the average number of photons of the coherent state $\ket{\alpha_l}$ is approximately given by
    \[\bar n \coloneqq {\abs{\alpha_l}}^2 \simeq \frac{R^{*2}}{4 \pzpf}.\]
    Depending on the value of the quantum scaling parameter $\pzpf$, many of the equilibria represented in Figure~\ref{fig:saddle_nodes} thus
    correspond to a very high average number of photons in the oscillator.
    Indeed for a ``nominal'' experimental value of $\pzpf = 0.2$, and for a reference value of $R = 3$, the corresponding mean number of photons amounts to
    \[\bar n = \frac{R^2}{4 \lambda^2} \simeq 56.\]
    Hence equilibria of the averaged model corresponding to very large $R^*$-values are not captured by the numerical simulations,
    due to a finite truncation of the Hilbert space.
    Depending on the value of $\pzpf$,
    we might not be able to capture such far-out equilibria of the classical system, as most of our simulations were performed using $300$ Fock states.
    Therefore, the number of Fock states was chosen as to capture the coherent states corresponding to the stable nodes of the averaged model that are \emph{closest} to the origin,
    for the smallest value of the quantum scaling parameter $\pzpf$ considered, namely $\pzpf = 0.2$.
    Such an analysis reveals that $150$ Fock states should suffice, and we see that the simulations have fully converged for $300$ Fock states,
    for any purple or black point in Figure~\ref{fig:phase_portraits}(bottom).
    We recall that the white points in Figure~\ref{fig:phase_portraits}(bottom) are not due a truncation of Fock space.
    All the relevant individual Floquet modes have fully converged with $300$ Fock states, but the specific mixture of Floquet modes in $\rho_\infty$
    is not uniquely defined for these points (see Remark~\ref{rem:non_convergence_new}).
\end{rem}

\bibliographystyle{plain}
\bibliography{refs.bib}